\documentclass[preprint,pra,showpacs]{revtex4}
\usepackage{amscd}
\usepackage{amsfonts}
\usepackage{amsmath}
\usepackage{amssymb}
\usepackage{color}
\usepackage{graphicx}
\usepackage{graphics}
\usepackage{hyperref}
\usepackage{wrapfig}
\usepackage[T1]{fontenc}
\usepackage[latin1]{inputenc}
\def\>{\rangle}
\def\<{\langle}
\def\n{\nonumber}

\begin{document}
\title{Electrostatic-field-induced dynamics in an ultrathin quantum well}
\author{Ilki Kim}
\email{hannibal.ikim@gmail.com} \affiliation{Department of Physics,
North Carolina A$\&$T State University, Greensboro, North Carolina
27411, USA}
\date{\today}
\begin{abstract}
We consider the time evolution of a particle subjected to both a
uniform electrostatic field $F$ and a one-dimensional delta-function
potential well. We derive the propagator ${\mathcal K}_F(x,t|x',0)$
of this system, directly leading to the wavefunction $\psi_F(x,t)$,
in which its essential ingredient ${\mathcal K}_F(0,t|0,0)$,
accounting for the ionization-recombination in the bound-continuum
transition, is exactly expressed in terms of the multiple
hypergeometric functions $F(z_1,z_2,\cdots,z_n)$. And then we obtain
the ingredient ${\mathcal K}_F(0,t|0,0)$ in an appropriate
approximation scheme, expressed in terms of the generalized
hypergeometric functions ${}_{p}\hspace*{-.04cm}F_q(z)$ being much
more transparent to physically interpret and much more accessible in
their numerical evaluation than the functions
$F(z_1,z_2,\cdots,z_n)$.
\end{abstract}
\maketitle
%
\section{Introduction}\label{sec:introduction}
The ionization of atoms in an external electric field is, as
well-known, one of the oldest subjects in quantum physics. As a
simple example of the ionization, hydrogen-like atoms in a uniform
electrostatic field have been extensively considered (see, e.g.,
\cite{YAM77,Raz03}). Here the background potential caused by the
field decreases without limit in one direction, and the electron
initially in the bound state will eventually tunnel through the
``barrier'' created by the field, so leading to ionization of the
atom. The tunneling rate for an {\em ensemble} of many independent
electrons has been calculated based on the exponential decay law
following from the statistical assumption that the tunneling rate is
proportional to the number of available atoms.

On the other hand, big experimental advances in the field of
nano-scale physics have highly enlarged a need for a detailed
understanding of the (time-dependent) tunneling process of {\em
individual} electrons subjected to an external field. In fact, the
nano-scale devices have been designed and fabricated, examples of
which are molecular switches \cite{JOA86} and resonant tunnel
junctions \cite{JON89}, etc. However, in exploring the time
evolution of the electron tunneling process leading to ionization,
we have a considerable mathematical difficulty that there are no
exactly solvable models for a transition from a bound state to the
continuum. Further, even obtaining the numerical solution, with high
accuracy, to this problem cannot be considered an easy task either,
especially in the strong-field limit where a highly oscillatory
behavior is found in the time evolution of the bound-continuum
transition.

The system under investigation in this paper is a particle subjected
to an attractive one-dimensional delta-function potential,
$-V_0\,\delta(x)$. We intend to study analytically the time
evolution of the particle when a uniform electrostatic field $F$ is
applied. The delta-function potential well (not in an external
field) has a single bound state. This aspect may rather simplify the
analytical study of the ionization. However, no exact solution
$\psi_F(x,t)$ to the time-dependent Schr\"{o}dinger equation of this
problem has been found in closed form (in terms of the actual
calculability to any sufficient degree of precision), even for such
a simple form of external field.

In fact, this model of the field-induced time-dependent ionization
has already been studied by some people. It was first discussed by
Geltman \cite{GEL77}. Later on, several different approaches to
obtaining the time-dependent wavefunction $\psi_F(x,t)$ have been
carried out \cite{LUD87,ELK88,SUS90,KLE94,ENG95,ROK00}. One of the
approaches mainly adopted so far is to turn the relevant
time-dependent Schr\"{o}dinger equation into the Lippmann-Schwinger
integral equation [cf. Eqs. (\ref{eq:lippmann_schwinger_eq1}) and
(\ref{eq:lippmann_schwinger_eq0})]. This integral equation is,
however, highly non-trivial to solve analytically and so has been
focused mostly upon its numerical solvability. In \cite{ELK88,ELB87}
by Elberfeld and Kleber, interestingly enough, the time-dependent
ionization probability has been investigated in the strong-field
limit based on the numerical analysis of the integral equation. They
also demonstrated numerically that the ionization probability
obtained from the simple exponential decay law may be considered a
fairly good approximation on the average to the exact result in the
strong-field limit although this approximation, by construction,
cannot account for the short-time ripples, observed in the exact
one, which result from the ionization-recombination process in the
bound-continuum transition \cite{ELK88,REI85}.

The goal of this paper lies in a systematic derivation of the
propagator ${\mathcal K}_F(x,t|x',0)$ of the system, directly
leading to $\psi_F(x,t) = \int_{-\infty}^{\infty} dx'\,{\mathcal
K}_F(x,t|x',0)\,\psi_0(x')$. As our central finding, its essential
constituent part ${\mathcal K}_F(0,t|0,0)$, accounting for the
ionization-recombination, is exactly expressed in terms of the
multiple hypergeometric functions $F(z_1,z_2,\cdots,z_n)$, and then
in an appropriate approximation scheme in terms of the generalized
hypergeometric functions ${}_{p}\hspace*{-.04cm}F_q(z)$ being much
more transparent to physically interpret and much more accessible in
their numerical evaluation than the functions
$F(z_1,z_2,\cdots,z_n)$. The general layout is the following. In
Sec. \ref{sec:review} we briefly review the well-known results of
both the delta-function potential problem without an external field
and the problem of a particle subjected to a uniform electrostatic
field but not bound by the ultrathin potential well. In doing so, we
also sophisticate the old results. In Sec. \ref{sec:propagator}, we
derive an explicit expression of the propagator in terms of the
multiple hypergeometric functions and then discuss its mathematical
complexity. Sec. \ref{sec:approximation} we obtain the propagator in
approximation, expressed in terms of the generalized hypergeometric
functions. Finally, we give the conclusion of this paper in Sec.
\ref{sec:conclusion}.

\section{General formulation}\label{sec:review}
The system under consideration is described by the Hamiltonian
\begin{equation}\label{eq:total_hamiltonian1}
    \hat{\mathcal H}_F\; =\; \frac{\hat{p}^2}{2 m}\, -\, V_0\,\delta(\hat{x})\, -\, \hat{x}\,F(t)\,,
\end{equation}
where an external field $F(t) = F \cdot \Theta(t)$, and $V_0 > 0$ is
a strength of the $\delta$-potential well. For the field-free case
($F=0$), it is well-known that the Hamiltonian $\hat{\mathcal H}_0$
has a single bound state \cite{GRI05}
\begin{equation}\label{eq:bound_state_H0}
    {\textstyle \psi_b(x)\; =\; \sqrt{B}\; e^{-B\,|x|}}
\end{equation}
with its eigen energy $E_b = -\hbar^2 B^2/2m$ where $B = m
V_0/\hbar^2$. All eigenstates and eigenvalues of $\hat{\mathcal
H}_0$ as well as the completeness of the eigenstates have been
discussed in detail in \cite{DAM75}. And for the potential-well-free
case ($V_0=0$), the eigenfunction of the Hamiltonian $\hat{H}_F =
\hat{p}^2/2m - \hat{x} F$ with (continuous) energy $E$ is given by
\cite{VAL04}
\begin{equation}\label{eq:eigenfunction_of_hamiltonian_f}
    \phi_E(x)\; =\; \left(\frac{4 m^2}{\hbar^4\,|F|}\right)^{1/6}\;
    \text{Ai}\left\{-\left(\frac{2 m F}{\hbar^2}\right)^{1/3}
    \left(x + \frac{E}{F}\right)\right\}
\end{equation}
in terms of the Airy function $\mbox{Ai}(z)$.

The method we employ here to obtain the propagator of the system
$\hat{\mathcal H}_F$ is to apply the relevant Lippmann-Schwinger
integral equation \cite{KLE94,WEI95}
\begin{equation}\label{eq:lippmann_schwinger_eq1}
    {\mathcal K}_F(x,t|x',0)\; =\; K_F(x,t|x',0)\,+\,\frac{i\,V_0}{\hbar} \int_0^t\, d\tau\, K_F(x,t|0,\tau) \cdot {\mathcal
    K}_F(0,\tau|x',0)\,,
\end{equation}
where the propagator ${\mathcal K}_F(x,t|x',0) := \<x|\exp(-i
t\,\hat{\mathcal H}_F/\hbar)|x'\>$. As well-known, the propagator
has the physical meaning of a (complex-valued)
transition-probability amplitude to get from the point $(x',0)$ to
the point $(x,t)$ \cite{GRO98}. Here $K_F(x,t|x',\tau)$ represents
the propagator relevant to the partial Hamiltonian $\hat{H}_F$ only,
explicitly given by \cite{ELK88}
\begin{equation}\label{eq:kernel1}
    K_F(x,t|x',\tau)\; =\;
    K_0\{x-x_c(t),\,t|x'-x_c(\tau),\,\tau\} \cdot
    e^{(1/i\hbar) \{S_c(t) - S_c(\tau)\}} \cdot e^{(i/\hbar) \{x\,p_c(t) -
    x' p_c(\tau)\}}\,,
\end{equation}
in which the propagator
\begin{equation}\label{eq:field_free_propagator}
    K_0(x,t|x',\tau)\; =\; \sqrt{\frac{m}{2 \pi i \hbar\,(t -
    \tau)}}\; \exp\left\{\frac{i}{\hbar}\,\frac{m}{2 (t - \tau)}\,(x -
    x')^2\right\}
\end{equation}
for a free particle subjected to $\hat{H}_0 = \hat{p}^2/2m$ only.
Also, $p_c(t) = \int_0^t d\tau\,F(\tau) = F\,t$ is a field-induced
classical impulse, leading to the corresponding field-induced
translation $x_c(t) = (1/m) \int_0^t d\tau\,p_c(\tau) = F t^2/2m$
and the field-induced action $S_c(t) = (1/2m) \int_0^t
d\tau\,p_c^2(\tau) = F^2 t^3/6m$. Eq.
(\ref{eq:lippmann_schwinger_eq1}) immediately gives rise to
\begin{equation}\label{eq:lippmann_schwinger_eq0}
    \psi_F(x,t)\; =\; \phi_F(x,t)\,+\,\frac{i\,V_0}{\hbar} \int_0^t d\tau\, K_F(x,t|0,\tau) \cdot
    \psi_F(0,\tau)\,,
\end{equation}
where the homogeneous solution
\begin{equation}\label{eq:homogeneous_solution1}
    \phi_F(x,t)\, =\, \int_{-\infty}^{\infty} dx'\,K_F(x,t|x',0)\,\psi_0(x')
\end{equation}
obviously represents free motion subjected to an external field $F$
only, while the second term on the right-hand side, by construction,
gives the influence of the residual zero-range potential. As an
example, for the initial bound state $\psi_0(x) = \psi_b(x)$, Eq.
(\ref{eq:homogeneous_solution1}) reduces to a closed expression
\cite{ELK88}
\begin{equation}\label{eq:homog_sol1}
    \phi_F(x,t)\; =\; \sqrt{B}\; e^{(i/\hbar)
    \left\{x\,p_c(t)-S_c(t)\right\}}\cdot\left\{M\left(x-x_c(t); -iB;
    \frac{\hbar t}{m}\right)\, +\, M\left(x_c(t)-x; -iB;
    \frac{\hbar t}{m}\right)\right\}
\end{equation}
in terms of the Moshinsky function \cite{MOS52}
\begin{equation}\label{eq:moshinsky_function}
    M(x;k;t)\; =\; \frac{1}{2}\, e^{i (k x - k^2
    t/2)}\; \mbox{erfc}\left(\frac{x - k t}{\sqrt{2 i t}}\right)
\end{equation}
where $\text{erfc}(z)$ is the complementary error function
\cite{ABS74}.

It is also instructive to point out that we can employee, as an
alternative to the interaction Hamiltonian $\hat{x}\,F$ in
(\ref{eq:total_hamiltonian1}) given in the scalar-potential gauge,
its counterpart $\hat{p}\,A$ in the vector-potential gauge, and the
system of interest is accordingly given by
\begin{equation}\label{eq:total_hamiltonian1-vector}
    \hat{\mathcal H}_A\; =\; \frac{1}{2 m} \left\{\hat{p}\,+\,p_c(t)\right\}^2\, -\,
V_0\,\delta(\hat{x})\,,
\end{equation}
where the vector potential $A$ corresponds to the field-induced
impulse $p_c(t)$. Then it can easily be verified that
\begin{subequations}
\begin{eqnarray}
    K_F(x,t|x',\tau) &=& \exp\left(\frac{i}{\hbar} \left\{x p_c(t)\,-\,x'
    p_c(\tau)\right\}\right) \cdot K_A(x,t|x',\tau)\label{eq:vector-potential-gauge-terms11}\\
    \psi_F(x,t) &=& \exp\left\{\frac{i}{\hbar}\,x \cdot p_c(t)\right\} \cdot \psi_A(x,t)\,.\label{eq:vector-potential-gauge-terms12}
\end{eqnarray}
\end{subequations}
Substituting (\ref{eq:vector-potential-gauge-terms11}) and
(\ref{eq:vector-potential-gauge-terms12}) into
(\ref{eq:lippmann_schwinger_eq1}) and
(\ref{eq:lippmann_schwinger_eq0}) respectively, we can
straightforwardly obtain the equivalent results in the
vector-potential gauge to what below follows in the scalar-potential
gauge (cf. for a detailed discussion of $\hat{x}\,F$ versus
$\hat{p}\,A$ gauge problem, see, e.g., \cite{SCH84}). Besides, our
formalism can apply to the system of a delta-potential barrier under
an electrostatic field
\begin{equation}\label{eq:total_hamiltonian1-1-0}
    \hat{\mathcal H}_F'\; =\; \frac{\hat{p}^2}{2 m}\, +\, V_0\,\delta(\hat{x})\, -\, \hat{x}\,F(t)
\end{equation}
as well, simply by replacing $V_0$ by $-V_0$ in Eq.
(\ref{eq:lippmann_schwinger_eq1}) or
(\ref{eq:lippmann_schwinger_eq0}) and then going forward
straightforwardly. In fact, the eigenstates and eigenvalues of the
Hamiltonian $\hat{{\mathcal H}}_0' = V_0\,\delta(\hat{x})$ are
identical to those of $\hat{{\mathcal H}}_0 =
-V_0\,\delta(\hat{x})$, except for the fact that $\hat{{\mathcal
H}}_0'$ does not have the bound state $\psi_b(x)$ as its eigenstate.

To develop the forthcoming discussions in a simplified fashion, let
us rescale coordinate and time as the dimensionless quantities,
$\tilde{x} = B x$ and $\tilde{t} = (\hbar B^2/m)\,t$, respectively
\cite{ELK88}, which is basically equivalent to setting $\hbar = m =
V_0 = 1$. Then the Schr\"{o}dinger equation for the Hamiltonian
$\hat{\mathcal{H}}_F$ easily reduces to
\begin{equation}\label{eq:schroedinger_eq2}
    \left(i \frac{\partial}{\partial \tilde{t}} + \frac{1}{2}
    \frac{\partial^2}{\partial {\tilde{x}}^2} + \delta(\tilde{x}) +
    \tilde{x} f\right) \psi_f(\tilde{x},\tilde{t})\, =\, 0\,,
\end{equation}
where the relative field strength $f = (m/\hbar^2 B^3)\,F$, and Eq.
(\ref{eq:lippmann_schwinger_eq1}) accordingly appears as
\begin{equation}\label{eq:lippmann_schwinger_eq2-0}
    {\mathcal K}_f(\tilde{x},\tilde{t}|\tilde{x}',0)\, =\, K_f(\tilde{x},\tilde{t}|\tilde{x}',0) + i \int_0^{\tilde{t}}
    d\tilde{\tau}\,K_f(\tilde{x},\tilde{t}|0,\tilde{\tau})\cdot{\mathcal K}_f(0,\tilde{\tau}|\tilde{x}',0)
\end{equation}
where the rescaled propagator
\begin{equation}\label{eq:kernel2}
    K_f(\tilde{x},\tilde{t}|\tilde{x}',\tilde{\tau}) =
    \sqrt{\frac{1}{2\pi i (\tilde{t}-\tilde{\tau})}}
    \exp\left\{\frac{i}{2\,(\tilde{t}-\tilde{\tau})} (\tilde{x}-\tilde{x}')^2 +
    \frac{i f}{2} (\tilde{x}+\tilde{x}') (\tilde{t}-\tilde{\tau}) -
    \frac{i f^2}{24}(\tilde{t}-\tilde{\tau})^3\right\}\,,
\end{equation}
which is identical to
$K_f(\tilde{x},\tilde{t}-\tilde{\tau}|\tilde{x}',0) =:
K_f(\tilde{x},\tilde{x}';\tilde{t}-\tilde{\tau})$. For the sake of
convenience we replace the notation $(\tilde{t},\tilde{x})$ by
$(t,x)$ for what follows.

In comparison, we also review briefly the exact results of the
field-free case ($f=0$). We first set $x=0$ in
(\ref{eq:lippmann_schwinger_eq2-0}), which then enables ${\mathcal
K}_0(0,t|x',0)$ on the left-hand side to immediately substitute for
${\mathcal K}_0(0,\tau|x',0)$ on the right-hand side. Making
iterations of the substitution, we can finally arrive at the
expression \cite{BAU85}
\begin{equation}\label{eq:field-free6}
    {\mathcal K}_0(x,t|x',0)\, =\, K_0(x,t|x',0)\,+\,\frac{1}{2}
    \sum_{n=1}^{\infty} (-2 z_2)^{n-1}\; \mbox{i}^{n-1} \mbox{erfc}(z_1)\,,
\end{equation}
where $z_1 = (|x|+|x'|)/\sqrt{2it}$ and $z_2 = t/i\sqrt{2it}$. Here
the index $n$ represents a number of the substitutions and so that
of interactions with the $\delta$-potential well, and $\mbox{i}^n
\mbox{erfc}(z)$ the repeated erfc integral [cf. $\mbox{i}^0
\mbox{erfc}(z) = \mbox{erfc}(z)$]. Using the identity \cite{ABS74}
\begin{equation}\label{eq:erfc-identity1}
    \left(\frac{d}{dz}\right)^n e^{z^2}\,\mbox{erfc}(z)\, =\,
    (-2)^n\,n!\,e^{z^2}\, \mbox{i}^n \mbox{erfc}(z)\,,
\end{equation}
we can easily obtain
\begin{equation}\label{eq:field-free7}
    \sum_{n=0}^{\infty} (-2 z_2)^n\, \mbox{i}^n \mbox{erfc}(z_1)\, =\,
    \exp\left(2 z_1 z_2 + z_2^2\right)\,\mbox{erfc}\left(z_1 +
    z_2\right)\,,
\end{equation}
which subsequently leads to the closed expression \cite{BAU85,ELB88}
\begin{equation}\label{eq:field-free8}
    {\mathcal K}_0(x,t|x',0)\, =\, K_0(x,t|x',0)\,+\,\frac{1}{2} M(|x|+|x'|; i; t)\,.
\end{equation}

Let us come back to the field-induced dynamics ($f \ne 0$) in order
to explore an explicit expression of ${\mathcal K}_f(x,t|x',0)$ as
the counterpart to the result in (\ref{eq:field-free8}). For doing
so, it may be useful to apply the Laplace transform ${\mathcal L}$
to Eq. (\ref{eq:lippmann_schwinger_eq2-0}) with respect to time; we
do this job first for $x=0$ with the help of the convolution
theorem, which will finally give rise to \cite{LUD87}
\begin{equation}\label{eq:general-propagator6}
    {\mathcal L}\{{\mathcal K}_f(x,t|x',0)\}(s)\, =:\, {\mathcal G}_f(x,x';s)\, =\, G_f(x,x';s)\, +\, i\,\frac{G_f(x,0;s)\,G_f(0,x';s)}{1 - i\,G_f(0,0;s)}\,,
\end{equation}
where the Laplace transform $\mathcal L\{K_f(x,x';t)\}(s) =:
G_f(x,x';s)$. From Eq. (\ref{eq:kernel2}) it can easily be shown as
well that $G_f(x,0;s) = G_f(0,x;s) =: G_f(x;s)$. To obtain an
explicit form of $G_f(x,x';s)$, we use the integral representation
\cite{ELK88}
\begin{equation}\label{eq:general-propagator8}
    \mbox{Ai}(\beta+|\alpha|)\, \mbox{Ci}(\beta-|\alpha|)\, =\,
    \frac{1}{2\pi} \int_0^{\infty} dt \sqrt{\frac{i}{\pi t}}\;
    \exp\left\{i\left(\frac{\alpha^2}{t} - \beta t -
    \frac{t^3}{12}\right)\right\}\,,
\end{equation}
where $\alpha$ and $\beta$ are real-valued, and the Airy function
$\mbox{Ci}(z) = \mbox{Bi}(z) + i \mbox{Ai}(z)$. This allows us to
have the Green function in the frequency domain \cite{LUD87}
\begin{equation}\label{eq:general-propagator9}
    G_f(x,x';s)|_{s\to -i\omega+0^+}\, :=\, \tilde{G}_f(x,x';\omega)\, =\,
    \left(\frac{4}{|f|}\right)^{\frac{1}{3}}\,\frac{\pi}{i}\;
    \mbox{Ai}(\beta+|\alpha|)\, \mbox{Ci}(\beta-|\alpha|)\,,
\end{equation}
where $|\alpha| \to (|f|/4)^{1/3}\,|x - x'|$ and $\beta \to
-(2\,|f|)^{-2/3} \{f\cdot(x + x') + 2 \omega\}$. Therefore, a closed
expression of the Green function $\tilde{{\mathcal
G}}_f(x,x';\omega)$, identical to ${\mathcal G}_f(x,x';s)$ in Eq.
(\ref{eq:general-propagator6}) with ${s \to -i \omega + 0^+}$,
immediately appears in terms of the Airy functions in
(\ref{eq:general-propagator9}). However, it is highly non-trivial to
directly carry out the inverse Fourier transform of the closed
expression of $\tilde{{\mathcal G}}_f(x,x';\omega)$ in the frequency
domain, even with the stationary phase approximation, in order to
obtain an explicit expression of ${\mathcal K}_f(x,t|x',0)$ in the
time domain \cite{KLE94}.

For a later purpose, we introduce here the time-dependent ionization
probability for the initial bound state $\psi_b(x)$, which is
defined as
\begin{equation}\label{eq:ionization_prob1}
    {\mathcal{P}}_f(t)\; :=\; 1 - \left|{\mathcal
    A}_{\psi_f}(t)\right|^2\,.
\end{equation}
Here the bound state amplitude
\begin{equation}\label{eq:bound_state_amplitude1}
    {\mathcal A}_{\psi_f}(t)\; =\; \int_{-\infty}^{\infty}
    dx\, \psi_b(x)\, \psi_f(x,t)\, =\, {\mathcal A}_{\phi_f}(t)\, +\, {\mathcal A}_{\delta}(t)\,,
\end{equation}
where, from Eq. (\ref{eq:lippmann_schwinger_eq0}),
\begin{subequations}
\begin{eqnarray}
    {\mathcal A}_{\phi_f}(t) &:=& \int_{-\infty}^{\infty} dx\, \psi_b(x)\, \phi_f(x,t)\label{q:second-term-of-ionization1}\\
    {\mathcal A}_{\delta}(t) &:=& i \int_{-\infty}^{\infty} dx\, \psi_b(x)\,
    \int_0^t d\tau\, K_f(x,t|0,\tau)\, \psi(0,\tau)\,.\label{q:second-term-of-ionization1-1}
\end{eqnarray}
\end{subequations}
Substituting (\ref{eq:homog_sol1}) into
(\ref{q:second-term-of-ionization1}), we can easily obtain a closed
form \cite{ELK88}
\begin{equation}\label{eq:homogeneous_ionization1}
    {\mathcal A}_{\phi_f}(t)\, =\,
    \frac{4}{f t}\, e^{-i f^2 t^3/6}\,\left\{\frac{M\left(\frac{f t^2}{2}; -i; t\right)}{2 i\,+\,f t}
    - \frac{M\left(-\frac{f t^2}{2}; -i; t\right)}{2 i\,-\,f t}\right\}\,,
\end{equation}
which reduces to unity at $t=0$, as required. Then it has been shown
\cite{KLE94,ELK88,LUD87} that in the weak-field limit $|f| \ll 1$,
the exponential decay law $|{\mathcal A}_{\psi_f}(t)|^2 \propto
e^{-\Gamma_f t}$ is a good approximation on the average. This
approximation can also be simulated by the ansatz in ``smooth'' form
\begin{equation}\label{eq:exponential_decay_ansatz}
    \psi_f(0,\tau)\; \propto\; e^{-i E \tau}\,,
\end{equation}
where the complex-valued energy $E\,=\,E_f - \frac{i}{2}\,\Gamma_f$
with $E_f\,=\,E_b + \Delta_f \in {\mathbb R}$ and the decay rate
$\Gamma_f \in {\mathbb R}$ \cite{KLE94,ELK88}. Here, $\Delta_f$ is
the level shift. The semiclassical value $\Gamma_{f,WKB} =
e^{-\frac{2}{3\,|f|}}$ is a good approximation of $\Gamma_f$ for
$|f| \lesssim 1$, and $\Delta_{f,WKB} = -\frac{5}{8} f^2$ is in
excellent agreement to $\Delta_f$ up to $|f| \lesssim 0.1$. However,
this exponential decay approximation, by construction, cannot
account for the ripples observed in the exact time evolution of the
ionization probability resulting from the ionization-recombination
process in the bound-continuum transition, explicitly demonstrated
in \cite{ELK88} from the numerical treatment of the
Lippmann-Schwinger equation.

\section{Derivation of an explicit expression of propagator}\label{sec:propagator}
To derive an explicit expression of the propagator, we make the same
iterations to Eq. (\ref{eq:lippmann_schwinger_eq2-0}) as those
applied above to the field-free case leading to
(\ref{eq:field-free6}), which then reveals that
\begin{eqnarray}\label{eq:propagator-expand1}
    \hspace{-1cm}&&\textstyle {\mathcal K}_f(x,t|x',0)\, =\, K_f(x,x';t)\,+\,i \int_0^t
    d\tau\,K_f(x,0;t-\tau)\,K_f(0,x';\tau)\,+\n\\
    \hspace{-1cm}&&\textstyle i^2 \int_0^t d\tau\,K_f(x,0;t-\tau) \int_0^{\tau} d\tau'\,K_f(0,0;\tau-\tau')\,K_f(0,x';\tau')\,+\,i^3
    \int_0^t d\tau\,K_f(x,0;t-\tau)\,\times\n\\
    \hspace{-1cm}&&\textstyle \int_0^{\tau} d\tau'\,K_f(0,0;\tau-\tau') \int_0^{\tau'}
    d\tau''\,K_f(0,0;\tau'-\tau'')\,K_f(0,x';\tau'')\,+\,\cdots\,.
\end{eqnarray}
To simplify the notation, let $K_f(x,0;t) = K_f(0,x;t) =: K_f(x;t)$
and then $K_f(0;t) =: K_f(t)$ from now on [cf. (\ref{eq:kernel2})].
For a later purpose, we rewrite Eq. (\ref{eq:propagator-expand1}) as
\begin{equation}\label{eq:propagator-expand-field1-0}
    {\mathcal K}_f(x,t|x',0)\, =\, K_f(x,x';t)\, +\, i \int_0^t
    d\tau\,K_f(x;t-\tau) \int_0^{\tau} d\tau'\,K_f(x';\tau-\tau')\cdot\Lambda(\tau')\,,
\end{equation}
where the partial integrand, being a function of time only,
\begin{eqnarray}\label{eq:propagator-field1-1-0}
    \Lambda(\tau') &:=& \delta(\tau')\, +\, i K_f(\tau')\, +\, i^2 \int_0^{\tau'}
    d\tau''\,K_f(\tau'-\tau'')\,K_f(\tau'')\, +\n\\
    && i^3 \int_0^{\tau'} d\tau''\,K_f(\tau'-\tau'') \int_0^{\tau''} d\tau'''\,K_f(\tau''-\tau''')\,K_f(\tau''')\,+\,\cdots\,.
\end{eqnarray}
Eq. (\ref{eq:propagator-expand-field1-0}) can easily be verified
with the help of the Laplace transform in
(\ref{eq:general-propagator6}), but without considering Eq.
(\ref{eq:general-propagator9}); the Laplace transform is
straightforwardly expanded as
\begin{equation}\label{eq:laplace_transform1}
    {\mathcal G}_f(x,x';s)\, =\, G_f(x,x';s)\,+\,i\,G_f(x;s)\, G_f(x';s) \sum_{n=0}^{\infty} \left\{i\,G_f(0;s)\right\}^n\,.
\end{equation}
And then we simply apply the inverse Laplace transform to this
expression term by term, which readily recovers the result in
(\ref{eq:propagator-expand-field1-0}). From this, it also follows
that
\begin{equation}\label{eq:propagator-expand-field1}
    \psi_f(x,t)\, =\, \phi_f(x,t)\, +\, i \int_0^t
    d\tau\,K_f(x;t-\tau) \int_0^{\tau} d\tau'\,\phi_f(0,\tau-\tau')\cdot\Lambda(\tau')\,.
\end{equation}
As seen, the finding of an explicit expression of the quantity
$\Lambda(\tau')$ is a critical step to deriving the wavefunction
$\psi_f(x,t)$ for any initial state $\psi_0(x)$ [cf. Eq.
(\ref{eq:lippmann_schwinger_eq0})].

Comments deserve here. First, the substitution of $x=x'=0$ into Eq.
(\ref{eq:propagator-expand-field1-0}) and then the comparison with
(\ref{eq:propagator-field1-1-0}) gives rise to a noteworthy relation
\begin{equation}\label{eq:lambda-propagator-0-0}
    {\mathcal K}_f(0,t|0,0)\, =\, i\,\delta(t) - i\,\Lambda(t)\,,
\end{equation}
which will be useful later. Secondly, the quantity $G_f(0;s)$ in
(\ref{eq:laplace_transform1}) designates the Green function for a
closed path with the initial and final point being $x=0$, and the
summation index $n$ represents how many times the closed path
interacts with the potential well located at $x=0$. Then the
quantity $\Lambda(\tau')$, exactly corresponding to this summation
in the time domain, is accordingly responsible for the
ionization-recombination resulting from the scattering from the
zero-range potential at $x=0$, particularly for $\psi_0(x) =
\psi_b(x)$, which was qualitatively explained in \cite{ELK88,REI85}
as follows: This initial bound state, explicitly given by
$\varphi_b(p) = \sqrt{2/\pi}/(p^2 + 1)$ in the momentum
representation, has a symmetric momentum distribution around $p = 0$
for the particles (in an ensemble represented by the bound state).
By applying the field, this symmetry breaks down in such a way that
the motion of the particles in one direction is accelerated and so
they will easily leave the potential well, simply toward the
(continuous) unbound states. On the other hand, the motion in the
other direction gets slowed down until the particles stop, and then
they reverse their direction of motion so that the particles again
approach the potential well at $x = 0$ where they cause a transient
maximum in the bound state probability. This process is repeated
until all particles will completely leave the potential well. Below
we will systematically derive the quantity $\Lambda(\tau')$, and so
${\mathcal K}_f(0,t|0,0)$, in closed form.

Let first $\Lambda_0(\tau') := \delta(\tau')$ and $\Lambda_1(\tau')
:= i K_f(\tau')$ in (\ref{eq:propagator-field1-1-0}). We consider
the next term $\Lambda_2(\tau') = - \mbox{I}_2(\tau')$, where the
integral
\begin{equation}\label{eq:propagator3}
    \mbox{I}_2(t)\, :=\, \int_0^t d\tau\,K_f(t-\tau)\, K_f(\tau)\, =\, \frac{1}{2\pi i} \int_0^t
    \frac{d\tau}{\sqrt{t-\tau}\,\sqrt{\tau}}\; \exp\left(-\frac{i f^2}{24} \left\{(t-\tau)^3 + \tau^3\right\}\right)
\end{equation}
[cf. (\ref{eq:kernel2})]. Let $\tau = t y$. Applying to Eq.
(\ref{eq:propagator3}) the Taylor series of the exponential function
and then the identity of beta function \cite{EXT76}
\begin{equation}\label{eq:identity_gamma0}
    \int_0^1 dy\; y^{\alpha-1}\, (1-y)^{\beta-1}\, =\,
    \frac{\Gamma(\alpha)\,\Gamma(\beta)}{\Gamma(\alpha+\beta)}\, =\, B(\alpha, \beta)
\end{equation}
where $\mbox{Re}(\alpha), \mbox{Re}(\beta) > 0$, it easily appears
that
\begin{equation}\label{eq:propagator20-0}
    \mbox{I}_2(t)\, =\, \frac{1}{2\pi i}\,\sum_{k,l=0}^{\infty} \frac{1}{k!\,l!}
    \left(\frac{f^2 t^3}{24 i}\right)^{k+l}\,
    \frac{\Gamma\left(3k+\frac{1}{2}\right)\,\Gamma\left(3l+\frac{1}{2}\right)}{\Gamma\left(3k+3l+1\right)}\,.
\end{equation}
With the help of the Gamma function identities, $\Gamma(3z) =
(2\pi)^{-1}\,3^{3z-1/2}\,\Gamma(z)\,\Gamma(z+1/3)\,\Gamma(z+2/3)$
and $\Gamma(1/6)\,\Gamma(1/2)\,\Gamma(5/6) = 2\pi^{3/2}$ and
$\Gamma(1/3)\,\Gamma(2/3) = 2\pi/3^{1/2}$ \cite{ABS74}, Eq.
(\ref{eq:propagator20-0}) reduces to a closed expression
\begin{equation}\label{eq:propagator20-1}
    \mbox{I}_2(t)\, =\, \frac{1}{2i} \times
    \left.F\begin{array}{rrr}
               -\hspace*{-.1cm}:&\hspace*{-.2cm}3\hspace*{-.01cm};&\hspace*{-.2cm}3\\[-.4cm]
               3\hspace*{-.1cm}:&\hspace*{-.2cm}-\hspace*{-.01cm};&\hspace*{-.2cm}-
           \end{array}\right.\left(\begin{array}{cccccc}
                                       -&:&\left[\frac{1}{6}, \frac{1}{2}, \frac{5}{6}\right]&;&\left[\frac{1}{6}, \frac{1}{2},
                                       \frac{5}{6}\right]&;\\
                                       \left[\frac{1}{3}, \frac{2}{3}, 1\right]&:&-&;&-&;
                                   \end{array}\,a, a
                             \right)
\end{equation}
with $a = f^2 t^3/24 i$ in terms of the generalized multiple
hypergeometric function \cite{SRI85,SRI87}
\begin{equation}\label{eq:multiple-hypergeometric0-1}
    \left.F\begin{array}{rrrr}
               p_0\hspace*{-.13cm}:&\hspace*{-.2cm}p_1\hspace*{-.05cm};&\hspace*{-.2cm}\cdots\hspace*{-.05cm};&\hspace*{-.2cm}p_n\\[-.4cm]
               q_0\hspace*{-.13cm}:&\hspace*{-.2cm}q_1\hspace*{-.05cm};&\hspace*{-.2cm}\cdots\hspace*{-.05cm};&\hspace*{-.2cm}q_n
           \end{array}\right.\left(\begin{array}{cccccccc}
                                       {\bf a}_0&:&{\bf a}_1&;&\cdots&;&{\bf a}_n&;\\
                                       {\bf b}_0&:&{\bf b}_1&;&\cdots&;&{\bf b}_n&;\\
                                   \end{array}\,z_1, \cdots, z_n
                             \right) := \sum_{k_1,\cdots,k_n = 0}^{\infty} \frac{({\bf a}_0)_{k_1+\cdots+k_n}}{({\bf b}_0)_{k_1+\cdots+k_n}}\,
                                        \prod_{j=1}^n \frac{({\bf a}_j)_{k_j}}{({\bf b}_j)_{k_j}}
                                        \frac{z_j^{k_j}}{k_j!}
\end{equation}
where ${\bf a}_j = (a_{j1}, \cdots, a_{j p_j})$ and ${\bf b}_j =
(b_{j1}, \cdots, b_{j q_j})$ with $j = 0, 1, \cdots, n$ are vectors
with dimensions $p_j$ and $q_j$, respectively. And $({\bf a}_j)_k :=
\prod_{l=1}^{p_j} (a_{jl})_k$ and $({\bf b}_j)_k :=
\prod_{l=1}^{q_j} (b_{jl})_k$, where the Pochhammer symbol
$(\lambda)_k = \Gamma(\lambda + k)/\Gamma(\lambda)$. The multiple
series in (\ref{eq:multiple-hypergeometric0-1}) absolutely converges
if $1 + q_0 + q_{j'} - p_0 - p_{j'} \geq 0$ for all $j' = 1, 2,
\cdots, n$. In fact, Eq. (\ref{eq:propagator20-1}) fulfills this
condition as $1 + 3 + 0 - 0 - 3 = 1\geq 0$ for $j' = 1, 2$.

Subsequently we consider the next term $\Lambda_3(\tau') =
-i\,\mbox{I}_3(\tau')$, where the integral
\begin{equation}\label{eq:propagator6}
    \mbox{I}_3(t)\, :=\, \int_0^t d\tau\,K_f(t-\tau) \int_0^{\tau} d\tau'\,K_f(\tau-\tau')\, K_f(\tau')\,.
\end{equation}
Similarly to $\mbox{I}_2(t)$, we can straightforwardly obtain
\begin{equation}\label{eq:propagator20}
    \mbox{I}_3(t)\, =\, \left(\frac{1}{2\pi i}\right)^{\frac{3}{2}}
    \sum_{k,l,n=0}^{\infty} \frac{1}{k!\,l!\,n!}
    \left(\frac{f^2}{24 i}\right)^{k+l+n} \int_0^t
    d\tau \int_0^{\tau} d\tau'\,(t-\tau)^{3k-\frac{1}{2}}\,(\tau-\tau')^{3l-\frac{1}{2}}\,(\tau')^{3n-\frac{1}{2}}\,.
\end{equation}
Let $\tau' = \tau y'$ and $\tau = t y$. We then use the identity
\cite{EXT76}
\begin{equation}\label{eq:identity_gamma1}
    \int_0^1 du \int_0^{1-u} dv\; u^{\alpha-1}\,v^{\beta-1}\,(1-u-v)^{\gamma-1}\,
    =\, \frac{\Gamma(\alpha)\,\Gamma(\beta)\,\Gamma(\gamma)}{\Gamma(\alpha+\beta+\gamma)}\,,
\end{equation}
where $\mbox{Re}(\alpha), \mbox{Re}(\beta), \mbox{Re}(\gamma) > 0$.
Eq. (\ref{eq:identity_gamma1}) is easily shown to be identical to
$\int_0^1 dw \int_0^w
du\,(1-w)^{\beta-1}\,(w-u)^{\gamma-1}\,u^{\alpha-1}$, where $w = 1 -
v$ and $\int_0^1 du \int_u^1 dw = \int_0^1 dw \int_0^w du$.
Consequently it appears that
\begin{equation}\label{eq:propagator21}
    \mbox{I}_3(t)\, =\, \left\{\frac{t}{(2\pi i)^3}\right\}^{\frac{1}{2}}
    \sum_{k,l,n=0}^{\infty} \frac{1}{k!\,l!\,n!}
    \left(\frac{f^2 t^3}{24 i}\right)^{k+l+n}\,
    \frac{\Gamma\left(3k+\frac{1}{2}\right)\,\Gamma\left(3l+\frac{1}{2}\right)\,\Gamma\left(3n+\frac{1}{2}\right)}{\Gamma\left(3k+3l+3n+\frac{3}{2}\right)}\,.
\end{equation}
With the aid of the technique already applied to
(\ref{eq:propagator20-0}), this reduces to a closed expression
\begin{equation}\label{eq:propagator22}
    \mbox{I}_3(t) = \left(\frac{t}{2\pi i^3}\right)^{\frac{1}{2}}
    \left.F\begin{array}{rrrr}
               -\hspace*{-.1cm}:&\hspace*{-.2cm}3\hspace*{-.01cm};&\hspace*{-.2cm}3\hspace*{-.01cm};&\hspace*{-.2cm}3\\[-.4cm]
               3\hspace*{-.1cm}:&\hspace*{-.2cm}-\hspace*{-.01cm};&\hspace*{-.2cm}-\hspace*{-.01cm};&\hspace*{-.2cm}-
           \end{array}\right.\left(\begin{array}{cccccccc}
                                       -&:&\left[\frac{1}{6}, \frac{1}{2}, \frac{5}{6}\right]&;&\left[\frac{1}{6}, \frac{1}{2},
                                       \frac{5}{6}\right]&;&\left[\frac{1}{6}, \frac{1}{2}, \frac{5}{6}\right]&;\\
                                       \left[\frac{1}{2}, \frac{5}{6}, \frac{7}{6}\right]&:&-&;&-&;&-&;
                                   \end{array}\,a, a, a
                             \right)
\end{equation}
with $a = f^2 t^3/24 i$. Along the same line, we can subsequently
obtain closed expressions of $\mbox{I}_4(t)$ and $\mbox{I}_5(t),
\cdots$, respectively, where $\Lambda_4(\tau') = \mbox{I}_4(\tau')$
and $\Lambda_5(\tau') = i\,\mbox{I}_5(\tau'), \cdots$ for Eq.
(\ref{eq:propagator-field1-1-0}).

Now let us generalize the above scenario to the quantity
$\Lambda_n(\tau') = i^n\,\mbox{I}_n(\tau')$, where the integral
\begin{equation}\label{eq:propagator23}
    \mbox{I}_n(t)\, :=\, \int_0^t d\tau_1\,K_f(t-\tau_1) \int_0^{\tau_1} d\tau_2\,K_f(\tau_1-\tau_2) \cdots \int_0^{\tau_{n-2}}
    d\tau_{n-1}\,K_f(\tau_{n-2}-\tau_{n-1})\, K_f(\tau_{n-1})\,.
\end{equation}
To do so, we apply the same technique with the aid of the identity
\cite{EXT76}
\begin{eqnarray}\label{eq:identity_gamma1-1}
    && \int_0^1 du_1 \int_0^{1-u_1} du_2\,\cdots \int_0^{1-u_1\cdots-u_{p-1}} du_p\;
    u_1^{\alpha_1-1}\cdots\,u_p^{\alpha_p-1}\,
    (1-u_1\cdots-u_p)^{\beta-1}\n\\
    &=& \frac{\Gamma(\alpha_1)\cdots\Gamma(\alpha_p)\,\Gamma(\beta)}{\Gamma(\alpha_1+\cdots+\alpha_p+\beta)}\,,
\end{eqnarray}
where $\mbox{Re}(\beta), \mbox{Re}(\alpha_j) > 0$ for $j =
1,2,\cdots, p$. This allows us to finally arrive at the expression
\begin{equation}\label{eq:propagator23-n}
    \mbox{I}_n(t)\, =\, \frac{1}{t} \left(\frac{t}{2\pi i}\right)^{\frac{n}{2}}
    \sum_{k_1,\cdots,k_n=0}^{\infty} \frac{\left(f^2 t^3/24
    i\right)^{k_1 + \cdots + k_n}}{k_1!\,\cdots\,k_n!}\;
    \frac{\Gamma\left(3k_1+\frac{1}{2}\right) \cdots \Gamma\left(3k_n+\frac{1}{2}\right)}{\Gamma\left(3k_1 + \cdots + 3k_n +
    \frac{N}{2}\right)}\,,
\end{equation}
which subsequently reduces to the closed expression
\begin{eqnarray}\label{eq:propagator23-2}
    \mbox{I}_n(t) &=&
    \left(\frac{t}{2i}\right)^{\frac{n}{2}} \frac{t^{-1}}{\Gamma\left(\frac{n}{2}\right)}\,
    \times\\
    && \left.F\begin{array}{rrcr}
                  -\hspace*{-.1cm}:&\hspace*{-.2cm}3\hspace*{-.01cm};&\hspace*{-.2cm}\cdots\hspace*{-.05cm};&\hspace*{-.2cm}3\\[-.4cm]
                  3\hspace*{-.1cm}:&\hspace*{-.2cm}-\hspace*{-.01cm};&\hspace*{-.2cm}\cdots\hspace*{-.05cm};&\hspace*{-.2cm}-
              \end{array}\right.\left(\begin{array}{cccccccc}
                                          -&:&\left[\frac{1}{6}, \frac{1}{2}, \frac{5}{6}\right]&;&\cdots&;&\left[\frac{1}{6},
                                          \frac{1}{2}, \frac{5}{6}\right]&;\\
                                          \left[\frac{n}{6}, \frac{n+2}{6}, \frac{n+4}{6}\right]&:&-_{\scriptscriptstyle{(1)}}&;&\cdots&;&-_{\scriptscriptstyle{(n)}}&;
                                      \end{array}\,\underbrace{a, \cdots, a}_{\scriptscriptstyle{n}}
                                \right)\,.\n
\end{eqnarray}
As a result, we find the exact expression
\begin{eqnarray}\label{eq:propagator-field1-1-0-0}
    \Lambda(\tau')\, =\, \delta(\tau')\, +\, i K_f(\tau')\, +\, \sum_{n=2}^{\infty} i^n\, \mbox{I}_n(\tau')
\end{eqnarray}
in terms of the (well-defined) multiple hypergeometric functions.
From (\ref{eq:lambda-propagator-0-0}), it easily follows as well
that ${\mathcal K}_f(0,t|0,0) = K_f(t) + \sum_{n=2}^{\infty}
i^{n-1}\,\mbox{I}_n(t)$. By substituting Eqs. (\ref{eq:kernel2}) and
(\ref{eq:propagator-field1-1-0-0}) into
(\ref{eq:propagator-expand-field1-0}), we can now obtain an explicit
expression of the propagator ${\mathcal K}_f(x,t|x',0)$ in terms of
the single double integral $\int_0^t d\tau \int_0^{\tau}
d\tau'\,\{\cdots\,\Lambda(\tau')\}$; this remaining double integral
cannot straightforwardly be evaluated indeed in terms of the
multiple hypergeometric functions in
(\ref{eq:multiple-hypergeometric0-1}). However, it is still
desirable to express Eq. (\ref{eq:propagator-field1-1-0-0}) (in good
approximation) in terms of the generalized hypergeometric functions
${}_{p}\hspace*{-.04cm}F_{q}(z)$ (with a single argument $z$)
\cite{EXT76} being much more transparent to physically interpret and
much more accessible in their numerical evaluation than the multiple
hypergeometric functions [cf. (\ref{eq:propagator-field1-1}) and
(\ref{eq:propagator-field2-1})].

In order to discuss the mathematical complexity of Eq.
(\ref{eq:propagator-field1-1-0-0}), it is also instructive to
formally express the quantity $\Lambda(\tau')$ in terms of a single
integral over the $n$-sphere solid angle $\Omega_n$ (e.g., a
$1$-sphere corresponding to a circle), in which the infinitesimal
element $d\Omega_n := (\sin\phi_n)^{n-1} \cdots\,\sin\phi_2\,d\phi_n
\cdots d\phi_2\,d\phi_1$, and the integral over the entire region is
explicitly given by $\oint d\Omega_n = \int_0^{\pi}
d\phi_n\,(\sin\phi_n)^{n-1} \cdots \int_0^{\pi} d\phi_2\,\sin\phi_2
\int_0^{2\pi} d\phi_1$ \cite{SOM58}. To this end, we first consider
the integral $\mbox{I}_2(t)$ in (\ref{eq:propagator3}). Let $\tau =
t\,(\sin\theta)^2$. This immediately allows us to have
\begin{equation}\label{eq:propagator4}
    \mbox{I}_2(t)\, =\, \frac{1}{\pi i} \int_0^{\frac{\pi}{2}} d\theta\, \exp\left(-\frac{i f^2 t^3}{24} \left\{(\cos\theta)^6 +
    (\sin\theta)^6\right\}\right)\,,
\end{equation}
which reduces to a closed form
\begin{equation}\label{eq:propagator5}
    \mbox{I}_2(t)\, =\, \frac{1}{2 i}\, \exp\left(-\frac{5 i f^2
    t^3}{192}\right)\,J_0\left(\frac{f^2 t^3}{64}\right)
\end{equation}
in terms of the Bessel function $J_0(z)$ [cf.
(\ref{eq:propagator20-1})]. Here we used $(\cos\theta)^6 +
(\sin\theta)^6 = 5/8 + (3/8)\cdot(\cos 4\theta)$ and then $J_0(z) =
(1/\pi) \int_0^{\pi} d\phi\,\cos(z \cos\phi)$ \cite{ABS74}.

The integral $\mbox{I}_3(t)$ is next under consideration. We
substitute Eqs. (\ref{eq:kernel2}) and (\ref{eq:propagator5}) into
(\ref{eq:propagator6}) and then carry out the same change of
integral variable as that used for (\ref{eq:propagator4}). Then it
appears that
\begin{equation}\label{eq:propagator7}
    \mbox{I}_3(t)\, =\, \left(\frac{t}{2\pi i^3}\right)^{\frac{1}{2}}
    \int_0^{\frac{\pi}{2}} d\theta\, \sin\theta\cdot\exp\left(-\frac{i f^2 t^3}{192}
    \left\{8\,(\cos\theta)^6 + 5 \,(\sin\theta)^6\right\}\right)\,J_0\left(\frac{f^2 t^3 (\sin\theta)^6}{64}\right)
\end{equation}
[cf. (\ref{eq:propagator17}) for exact evaluation of this integral].
We next pay an explicit attention to
\begin{equation}\label{eq:propagator-7-0}
    \mbox{I}_4(t)\, =\,
    \int_0^t d\tau_1\,K_f(t-\tau_1) \int_0^{\tau_1} d\tau_2\,K_f(\tau_1-\tau_2) \int_0^{\tau_2}
    d\tau_3\,K_f(\tau_2-\tau_3)\, K_f(\tau_3)
\end{equation}
[cf. Eq. (\ref{eq:propagator23}) with $n=4$]. Using $\int_0^t d\tau
\int_0^{\tau} d\tau' = \int_0^t d\tau' \int_{\tau'}^t d\tau$ and
then introducing $u := \tau - \tau'$, the first two-integral part of
(\ref{eq:propagator-7-0}) is easily transformed into $\int_0^t
d\tau' \int_0^{t-\tau'} du\,K_f(t-\tau'-u)\,K_f(u)$. This, with Eqs.
(\ref{eq:propagator3}) and (\ref{eq:propagator5}), then allows us to
have
\begin{equation}\label{eq:propagator25}
    \mbox{I}_4(t)\, =\, \left(\frac{1}{2i}\right)^2
    \int_0^t d\tau\,\exp\left(\frac{-5 i f^2 \{(t-\tau)^3 +
    \tau^3\}}{192}\right)\,J_0\left(\frac{f^2
    (t-\tau)^3}{64}\right)\,J_0\left(\frac{f^2 \tau^3}{64}\right)\,.
\end{equation}
Applying again the very technique already used for $\mbox{I}_2(t)$
and $\mbox{I}_3(t)$, we can finally obtain
\begin{eqnarray}\label{eq:propagator-5}
    \mbox{I}_4(t) &=& \frac{t}{(2i)^2} \int_0^{\frac{\pi}{2}} d\theta\, (\sin
    2\theta)\, \exp\left(\frac{-5 i f^2 t^3}{192} \{(\cos\theta)^6 +
    (\sin\theta)^6\}\right)\,\times\n\\
    && J_0\left(\frac{f^2 t^3
    (\cos\theta)^6}{64}\right)\,J_0\left(\frac{f^2 t^3 (\sin\theta)^6}{64}\right)
\end{eqnarray}
[cf. (\ref{eq:propagator-III-8})]. Similarly, we can also find that
\begin{eqnarray}\label{eq:propagator-61}
    \mbox{I}_5(t) &=& \textstyle\left(\frac{t^3}{8 \pi i^5}\right)^{\frac{1}{2}}
    \int_1 d\Omega_2\,(\sin\theta)^2\, (\sin 2\varphi)\, J_0\left(\frac{f^2 t^3
    (\sin\theta)^6\,(\cos\varphi)^6}{64}\right)\, J_0\left(\frac{f^2 t^3
    (\sin\theta)^6\,(\sin\varphi)^6}{64}\right)\,\times\n\\
    && \exp\left\{\frac{-i f^2 t^3}{192} \left(8\,(\cos\theta)^6 +
    5\,(\sin\theta)^6 \left\{(\cos\varphi)^6 +
    (\sin\varphi)^6\right\}\right)\right\}\,,
\end{eqnarray}
where $\int_1 d\Omega_2 \cdots = \int_0^{\pi/2} \int_0^{\pi/2}
\sin\theta\,d\theta\,d\varphi \cdots$ represents an integral over
the surface of a unit $2$-sphere (with radius $r=1$) covering the
first octant only. Along the same line, we continue to be able to
obtain the corresponding expressions of $\mbox{I}_6(t)$ and
$\mbox{I}_7(t), \cdots$, respectively. In doing this job, it can
easily be induced that
\begin{subequations}
\begin{eqnarray}
    \hspace*{-1cm}\mbox{I}_{2m+1}(t) &=& \left(\frac{2 t}{\pi i}\right)^{\frac{1}{2}} \int_0^{\frac{\pi}{2}}
    d\phi_m\, \sin\phi_m\cdot\exp\left\{\frac{-i f^2 t^3
    \left(\cos\phi_m\right)^6}{24}\right\}\cdot\mbox{I}_{2m}\left(t\,(\sin\phi_m)^2\right)\label{eq:iterated1}\\
    \hspace*{-1cm}\mbox{I}_{2m+2}(t) &=& \frac{t}{2i} \int_0^{\frac{\pi}{2}}
    d\phi_m\, (\sin 2\phi_m)\, \exp\left\{\frac{-5 i f^2
    t^3 \left(\cos\phi_m\right)^6}{192}\right\}\,
    J_0\left(\frac{f^2 t^3 (\cos\phi_m)^6}{64}\right)\,\times\n\\
    && \mbox{I}_{2m}\left(t\,(\sin\phi_m)^2\right)\,,\label{eq:iterated2}
\end{eqnarray}
\end{subequations}
where $m = 1, 2, \cdots$.

Now let us generalize the above scenario to the integral
$\mbox{I}_n(t)$. Applying the iterations to Eqs.
(\ref{eq:propagator5}), (\ref{eq:iterated1}) and
(\ref{eq:iterated2}) allows us to finally obtain the following
irreducible expressions in terms of an integral over the
higher-dimensional solid angle as
\begin{subequations}
\begin{eqnarray}
    \mbox{I}_{2m+1}(t) &=&
    \left(\frac{1}{2\pi i t}\right)^{\frac{1}{2}} \left(\frac{t}{i}\right)^m
    \int_1 d\Omega_m\, \exp\left\{\frac{-i f^2 t^3
    \left(\cos\phi_m\right)^6}{24}\right\}\,\times\n\\
    && \prod_{k=1}^m \left\{(\sin\phi_k)^k\,(\cos\phi_{k-1})\;
    \exp\left(\frac{-5 i f^2 t^3}{192}\,(\cos\phi_{m-k})^6
    \prod_{l=m-k+1}^m
    (\sin\phi_l)^6\right)\,\times\right.\n\\
    && \left.J_0\left(\frac{f^2 t^3}{64}\,(\cos\phi_{m-k})^6
    \prod_{l=m-k+1}^m
    (\sin\phi_l)^6\right)\right\}\label{eq:iterated1-1}
\end{eqnarray}
and
\begin{eqnarray}
    \hspace*{-.3cm}\mbox{I}_{2m+2}(t) &=& {\textstyle\frac{1}{2 i}
    \left(\frac{t}{i}\right)^m
    \int_1 \frac{d\Omega_m}{\left(\sin\phi_{m+1}\right)^{m+1}}}
    \prod_{k=1}^{m+1} \textstyle\left\{(\sin\phi_k)^k\,(\cos\phi_{k-1})\,
    \exp\left(-\frac{5 i f^2 t^3}{192}\,\left(\frac{\cos\phi_{m-k+1}}{\sin\phi_{m+1}}\right)^6\,\times\right.\right.\n\\
    \hspace*{-.3cm}&& \left.\left.\prod_{l=m-k+2}^{m+1}
    \left(\sin\phi_l\right)^6\right)\cdot
    J_0\left(\frac{f^2 t^3}{64}\,\left(\frac{\cos\phi_{m-k+1}}{\sin\phi_{m+1}}\right)^6
    \prod_{l=m-k+2}^{m+1}
    (\sin\phi_l)^6\right)\right\}\,.\label{eq:iterated2-1}
\end{eqnarray}
\end{subequations}
Here the angle $\phi_0 := 0$, and $\int_1 d\Omega_m \cdots$
represents an integral over the first section only, corresponding to
$0 \leq \phi_1, \cdots, \phi_m \leq \pi/2$. Now we easily see that
with the field strength $f \to \infty$, the integrals $\mbox{I}_n(t)
\to 0$ for $n \geq 2$, due to the fact that $J_0(z) \to 0$ with $z
\to \infty$. As a result, the quantity $\Lambda(\tau')$ in
(\ref{eq:propagator-field1-1-0-0}) can be rewritten in terms of the
integrals over the higher-dimensional solid angle. This result can
also be interpreted in such a way that Eq.
(\ref{eq:propagator-field1-1-0}) for $\Lambda(\tau')$, given in form
of the {\em time-ordered} integrals $\int_0^{\tau'} d\tau''
\int_0^{\tau''} d\tau''' \cdots$ and so non-trivial to directly
evaluate, is transformed, for an arbitrary time $\tau'$, into the
integrals of some geometric pattern over the (time-independent)
surface of a unit $n$-sphere, so being more accessible in numerical
evaluation; in fact, we see from (\ref{eq:iterated1-1}) and
(\ref{eq:iterated2-1}) that the quantity $\mbox{I}_n(t)$ is given by
an irreducible $[(n-1)/2]$-dimensional integral, rather than an
$(n-1)$-dimensional integral in (\ref{eq:propagator-field1-1-0}),
where the symbol $[y]$ is the greatest integer less than or equal to
$y$. Due to this irreducible high dimensionality, it is, apparently,
highly non-trivial, though, to exactly evaluate the (complicated)
integral $\mbox{I}_n(t)$ for $n$ being large enough in terms of the
hypergeometric functions ${}_{p}\hspace*{-.04cm}F_{q}(z)$ simply
with a single argument $z$. Below we will accordingly explore an
approximation scheme in which the quantity $\Lambda(\tau')$ can be
expressed in terms of the functions ${}_{p}\hspace*{-.04cm}F_{q}(z)$
in reasonably simple form.

\section{Approximation of Propagator based on the Partial Wave Expansion}\label{sec:approximation}
%
We already have the closed expression of the integral
$\mbox{I}_2(t)$ in terms of the Bessel function in
(\ref{eq:propagator5}). We next intend to evaluate the integral
$\mbox{I}_3(t)$ in (\ref{eq:propagator7}) in terms of some
hypergeometric function. Substituting first the identity
\cite{ABS74}
\begin{equation}\label{eq:bessel-function1}
    J_{\nu}(z)\, =\,
    \frac{(z/2)^{\nu}\,e^{-i z}}{\Gamma(\nu+1)}\,
    {}_{1}\hspace*{-.07cm}F_1\left(\nu+\frac{1}{2}; 2\nu+1;
    2iz\right)
\end{equation}
with $\nu = 0$ into (\ref{eq:propagator7}), we can straightforwardly
obtain
\begin{equation}\label{eq:propagator8}
    \mbox{I}_3(t)\, =\, \left(\frac{t}{2\pi^2\,i^3}\right)^{\frac{1}{2}} \exp\left(-\frac{5 i f^2 t^3}{192}\right)
    \sum_{n=0}^{\infty} \frac{\Gamma(\frac{1}{2} +
    n)}{\Gamma(1+n)} \frac{1}{n!} \left(\frac{i f^2 t^3}{32}\right)^n \times
    \mbox{I}_{31}(t)\,,
\end{equation}
where
\begin{equation}\label{eq:propagator9}
    \mbox{I}_{31}(t)\, :=\, \int_0^{\frac{\pi}{2}} d\theta\,(\sin\theta)^{6n+1}\,\exp\left(-\frac{i f^2 t^3}{64} \cos 4\theta\right)\,.
\end{equation}
Here the confluent hypergeometric function
${}_{1}\hspace*{-.07cm}F_1(a; b; z) = \{\Gamma(b)/\Gamma(a)\}
\sum_{n=0}^{\infty} \{\Gamma(a+n)/\Gamma(b+n)\}\,z^n/n!$.

Now we focus on explicit evaluation of the integral
$\mbox{I}_{31}(t)$. To do so, we first substitute into
(\ref{eq:propagator9}) the expansion formula for a plane wave
\cite{ABS74}
\begin{equation}\label{eq:identity_bessel1}
    \exp\left(i a \cos\phi\right)\, =\, \sum_{l=-\infty}^{\infty} i^l
    J_l(a)\,\cos(l\phi)
\end{equation}
with $\phi \to 4\theta$ and $a \to -f^2 t^3/64$, where the Bessel
functions $J_l(z)$. As seen, each $l$ of the harmonic partial waves
is then time-independent. Subsequently, with the help of another sum
rule $\cos(nz) = \sum_{k=0}^n \binom{n}{k}\,(\cos z)^k\,(\sin
z)^{n-k}\,\cos\{(n-k)\,\pi/2\}$ with $n = 4l$ followed by the beta
function $B(w,z) = 2 \int_0^{\pi/2}
d\phi\,(\sin\phi)^{2w-1}\,(\cos\phi)^{2z-1}$ [cf.
(\ref{eq:identity_gamma0})], we can easily obtain the integral
representation \cite{KIM12}
\begin{equation}\label{eq:propagator14}
    \int_0^{\frac{\pi}{2}} d\theta\,(\sin\theta)^p\, (\cos 4l\theta)\, =\,
    \frac{\pi\; \Gamma(p+1)}{2^{p+1}\; \Gamma\left(1 + 2l + \frac{p}{2}\right)\, \Gamma\left(1 - 2l +
    \frac{p}{2}\right)}\,,
\end{equation}
in which $p \to 6n+1$. In doing so, we also used
$\Gamma(z)\,\Gamma(1-z) = \pi\,\mbox{csc}(\pi z)$ and $\Gamma(2z) =
(2\pi)^{-1/2}\,2^{2z-1/2}\,\Gamma(z)\,\Gamma(z+1/2)$ \cite{ABS74}.
Eqs. (\ref{eq:identity_bessel1}) and (\ref{eq:propagator14}) then
allow us to have an evaluation of the integral $\mbox{I}_{31}(t)$
and so
\begin{equation}\label{eq:propagator16}
    \mbox{I}_3(t)\, =\, {\textstyle\left(\frac{t}{2^5\,i^3}\right)^{\frac{1}{2}}\,\exp\left(-\frac{5 i f^2 t^3}{3\cdot 2^6}\right)}
    \sum_{n=0}^{\infty} {\textstyle\frac{\Gamma\left(\frac{1}{2} + n\right)\cdot (6n + 1)!}{\left(n!\right)^2}
    \left(\frac{i f^2 t^3}{2^{11}}\right)^n} \sum_{l=-\infty}^{\infty}
    {\textstyle\frac{i^l\,J_l\left(-\frac{f^2 t^3}{2^6}\right)}{\Gamma\left(\frac{3}{2} + 3n + 2l\right)\,\Gamma\left(\frac{3}{2} + 3n -
    2l\right)}}\,.
\end{equation}
This subsequently reduces to a compact expression
\begin{equation}\label{eq:propagator17}
    \mbox{I}_3(t)\, =\, \left(\frac{t}{2\pi i^3}\right)^{1/2}
    \exp\left(-\frac{5 i f^2 t^3}{192}\right)
    \sum_{l=-\infty}^{\infty} \frac{i^{-l}}{(1-16\,l^2)}\; J_l\left(\frac{f^2 t^3}{64}\right)\; F_l^{(6)}\left(\frac{i f^2 t^3}{32}\right)
\end{equation}
in terms of the generalized hypergeometric functions
\begin{equation}\label{eq:generalized-hypergeometric1}
    F_l^{(6)}(z)\, :=\,
    {}_{6}\hspace*{-.07cm}F_6\left(\left[\frac{1}{3}, \frac{1}{2}, \frac{1}{2}, \frac{2}{3}, \frac{5}{6}, \frac{7}{6}\right];
    \left[\frac{1}{2}-\frac{2l}{3}, \frac{1}{2}+\frac{2l}{3}, \frac{5}{6}-\frac{2l}{3}, \frac{5}{6}+\frac{2l}{3}, \frac{7}{6}-\frac{2l}{3},
    \frac{7}{6}+\frac{2l}{3}\right]; z\right)\,,
\end{equation}
which satisfies
\begin{equation}\label{eq:propagator18}
    \sum_{n=0}^{\infty} \frac{\Gamma\left(\frac{1}{2} + n\right)\cdot (6n + 1)!}{\Gamma(1 +
    n)\,\Gamma\left(\frac{3}{2} + 2l + 3n\right)\,\Gamma\left(\frac{3}{2} - 2l + 3n\right)}\,\frac{z^n}{n!}\, =\,
    \frac{4\cdot\pi^{-1/2}}{1 - 16\,l^2}\; F_l^{(6)}\left(2^6 z\right)\,.
\end{equation}
Here we also used $J_l(-z) = (-)^l J_l(z)$.

Comments deserve here. First, Eq. (\ref{eq:propagator17}) can be
rewritten as
\begin{eqnarray}\label{eq:propagator19}
    \mbox{I}_3(t) &=& \left(\frac{t}{2\pi i^3}\right)^{\frac{1}{2}}
    \exp\left(-\frac{5 i f^2 t^3}{192}\right)
    \left\{J_0\left(\frac{f^2 t^3}{64}\right)\; {}_{2}\hspace*{-.07cm}F_2\left(\left[\frac{1}{3}, \frac{2}{3}\right];
    \left[\frac{5}{6}, \frac{7}{6}\right]; \frac{i f^2
    t^3}{32}\right)\, +\right.\n\\
    && \left.\sum_{l=1}^{\infty} \frac{2\,i^{-l}}{1-16\,l^2}\; J_l\left(\frac{f^2 t^3}{64}\right)\; F_l^{(6)}\left(\frac{i f^2
    t^3}{32}\right)\right\}
\end{eqnarray}
[cf. (\ref{eq:propagator22}) and (\ref{eq:bessel-function1})]. Now
it is explicitly shown that the quantity $\mbox{I}_2(t)$ in
(\ref{eq:propagator5}) is ``modulated'' by
${}_{2}\hspace*{-.07cm}F_2(\cdots)$ yielding in
(\ref{eq:propagator19}) the partial wave $l=0$ as the leading term
of $\mbox{I}_3(t)$, surrounded by the additional partial waves $l\ne
0$ ``modulated'' by $F_l^{(6)}(\cdots)$. Secondly, we may
accordingly take the leading term only as a satisfactory
approximation of $\mbox{I}_3(t)$; in the weak-field regime ($f \ll
1$) leading to $J_l(b) \to 0$ for $l \ne 0$ with $b = f^2 t^3/64$,
we easily see the validity of this approximation. It also applies
sufficiently to the strong-field regime ($f \gg 1$), in which due to
the asymptotic behavior $J_l(b) \approx \{2/(\pi b)\}^{1/2} \cos(b -
l\pi/2 - \pi/4)$ \cite{ABS74}, the magnitude of all terms with $l
\ne 0$ cannot be non-negligible enough. In fact, we have
$|\mbox{I}_3(t)| \ll 1$ anyway in this regime, as already pointed
out after Eq. (\ref{eq:iterated2-1}). Also, the strong-field regime
corresponds to the semiclassical limit $\hbar \to 0$ since the
argument $b$, expressed in a dimensionless unit, exactly corresponds
to $F^2 t^3/(64 \hbar m)$ in the actual physical unit [cf.
(\ref{eq:schroedinger_eq2})]. Therefore the approximation with $l=0$
alone may also be considered an effective semiclassical treatment of
$\mbox{I}_3(t)$.

Next we explore a closed expression of $\mbox{I}_4(t)$ in
approximation. The substitution of the leading term of
$\mbox{I}_3(t)$ into Eq. (\ref{eq:propagator-7-0}) allows us to
straightforwardly obtain
\begin{equation}\label{eq:propagator-III-1}
    \mbox{I}_4(t)\,
    \approx\, -\frac{1}{8\pi} \sum_{n=0}^{\infty}
    \frac{\Gamma\left(\frac{1}{2}+n\right)\; \Gamma(2+6n)}{\left\{\Gamma(1+n)\; \Gamma\left(\frac{3}{2}+3n\right)\right\}^2}
    \left(\frac{i f^2}{2^{11}}\right)^n \sum_{k=0}^{\infty}
    \frac{\Gamma(\frac{1}{2}+k)}{\left\{\Gamma(1+k)\right\}^2} \left(\frac{i f^2}{2^5}\right)^k \times \mbox{I}_{41}(t)\,,
\end{equation}
where
\begin{eqnarray}\label{eq:propagator-III-2}
    \mbox{I}_{41}(t) &:=& \int_0^t d\tau\, \frac{\tau^{3 (n+k) +
    1/2}}{\sqrt{t-\tau}}\, \exp\left(-\frac{i f^2}{24} \{(t-\tau)^3 +
    \tau^3\}\right)\n\\
    &=& 2\, t^{3(n+k)+1}\, \exp\left(-\frac{5 i f^2 t^3}{192}\right)
    \int_0^{\frac{\pi}{2}} d\theta\; (\sin\theta)^{6(n+k)+2}\,
    \exp\left(-\frac{i f^2 t^3}{64} \cos 4\theta\right)\,.
\end{eqnarray}
Here we applied the same technique as that used for
(\ref{eq:propagator8})-(\ref{eq:propagator9}). Subsequently we again
apply to Eq. (\ref{eq:propagator-III-2}) both the partial wave
expansion in (\ref{eq:identity_bessel1}), followed by the selection
of $l=0$ alone, and Eq. (\ref{eq:propagator14}) with $p \to
6(n+k)+2$. This immediately gives rise to
\begin{equation}\label{eq:propagator-III-3}
    \mbox{I}_{41}(t)\, \approx\,
    \pi \left(\frac{t}{4}\right)^{3(n+k)+1} \exp\left(-\frac{5 i f^2 t^3}{192}\right)
    J_0\left(\frac{f^2 t^3}{64}\right)
    \frac{\Gamma\{6(n+k)+3\}}{(\Gamma\{3(n+k)+2\})^2}\,.
\end{equation}
Substituting this into (\ref{eq:propagator-III-1}), we can next
obtain straightforwardly
\begin{equation}\label{eq:propagator-III-4}
    \mbox{I}_4(t)\, \approx\, -\frac{t}{32}\, \exp\left(-\frac{5 i f^2
    t^3}{192}\right)\, J_0\left(\frac{f^2 t^3}{64}\right) \times
    \mbox{I}_{42}(t)\,,
\end{equation}
where
\begin{equation}\label{eq:propagator-III-5}
    \mbox{I}_{42}(t)\, :=\, \sum_{n=0}^{\infty}
    \frac{\Gamma\left(\frac{1}{2}+n\right)\,\Gamma(2+6n)}{\left\{\Gamma\left(\frac{3}{2}+3n\right)\cdot n!\right\}^2}
    \left(\frac{i f^2 t^3}{2^{17}}\right)^n \sum_{k=0}^{\infty}
    \frac{\Gamma(\frac{1}{2}+k)\,\Gamma\{3+6(n+k)\}}{\left(\Gamma\{2+3(n+k)\}\cdot k!\right)^2}
    \left(\frac{i f^2 t^3}{2^{11}}\right)^k\,.
\end{equation}
Let $n+k = r$. Then, this can easily be rewritten as
\begin{equation}\label{eq:propagator-III-6}
    \mbox{I}_{42}(t)\, =\, \sum_{r=0}^{\infty}
    \frac{\Gamma(3+6r)}{\{\Gamma(2+3r)\}^2} \left(\frac{i f^2
    t^3}{2^{11}}\right)^r \sum_{n=0}^r
    \frac{\Gamma\left(\frac{1}{2}+n\right)\,\Gamma(2+6n)\,\Gamma\left(\frac{1}{2}+r-n\right)}{\left\{\Gamma\left(\frac{3}{2}+3n\right)\,
    \Gamma(1+r-n)\cdot n!\right\}^2} \left(\frac{1}{2^6}\right)^n\,.
\end{equation}
Here the second summation over the index $n$ precisely simplifies to
$4\cdot\Gamma\left(\frac{1}{2}+r\right)/\{\sqrt{\pi}\,(r!)^2\}\cdot
{}_{4}\hspace*{-.07cm}F_3\left(\left[\frac{1}{3},\frac{2}{3},-r,-r\right];
\left[\frac{5}{6},\frac{7}{6},\frac{1}{2}-r\right]; -1\right)$ and
can be approximated with satisfactory precision to its leading term
$n=0$ alone, due to the fact that the summand $(1/2^6)^n \to 0$ for
all $n \geq 1$ (cf. Fig. \ref{fig:fig1}). From this, Eq.
(\ref{eq:propagator-III-6}) reduces to
\begin{equation}\label{eq:propagator-III-7}
    \mbox{I}_{42}(t)\, \approx\, \frac{4}{\sqrt{\pi}} \sum_{r=0}^{\infty}
    \frac{\Gamma\left(\frac{1}{2}+r\right)\,\Gamma(3+6r)}{\left\{\Gamma(2+3r)\cdot r!\right\}^2}
    \left(\frac{i f^2 t^3}{2^{11}}\right)^r\, =\,
    8\, {}_{4}\hspace*{-.07cm}F_4\left(\left[\frac{1}{2},\frac{1}{2},\frac{5}{6},\frac{7}{6}\right];
    \left[\frac{2}{3},1,1,\frac{4}{3}\right]; \frac{i
    f^2t^3}{32}\right)
\end{equation}
which immediately gives rise to the expression
\begin{equation}\label{eq:propagator-III-8}
    \mbox{I}_4(t)\, \approx\, -\frac{t}{4}\,\exp\left(-\frac{5 i f^2 t^3}{192}\right) J_0\left(\frac{f^2
    t^3}{64}\right)\, {}_{4}\hspace*{-.07cm}F_4\left(\left[\frac{1}{2},\frac{1}{2},\frac{5}{6},\frac{7}{6}\right];
    \left[\frac{2}{3},1,1,\frac{4}{3}\right]; \frac{i
    f^2t^3}{32}\right)
\end{equation}
in terms of the generalized hypergeometric function. As demonstrated
in (\ref{eq:propagator-III-1})-(\ref{eq:propagator-III-7}), without
this leading-term approximation the quantity $\mbox{I}_4(t)$ should
be expressed in terms of a lengthy multiple sum, which is not
transparent to physically interpret and not desirable for numerical
evaluation.

Next the quantity $\mbox{I}_5(t)$ is under consideration. We exactly
follow the technique leading to Eq. (\ref{eq:propagator-III-8}) for
$\mbox{I}_4(t)$; we first substitute this previous result with
(\ref{eq:propagator-III-7}) into (\ref{eq:propagator23}) with $n=5$
and then carry out the same approximation, followed by applying Eq.
(\ref{eq:propagator14}) with $p \to 6(n+k)+3$. From this, we can
arrive at the rather lengthy expression
\begin{equation}\label{eq:propagator-IV-4-2}
    \mbox{I}_5(t)\, \approx\, -\frac{t^{3/2}}{64} \sqrt{\frac{1}{2\pi i}}\, \exp\left(-\frac{5 i f^2
    t^3}{192}\right) J_0\left(\frac{f^2 t^3}{64}\right) \times \mbox{I}_{51}(t)\,,
\end{equation}
where
\begin{equation}\label{eq:propagator-IV-4-3}
    \mbox{I}_{51}(t)\, :=\, \sum_{n=0}^{\infty}
    \frac{\Gamma\left(\frac{1}{2}+n\right)\,\Gamma(3+6n)}{\left\{\Gamma\left(2+3n\right)\cdot n!\right\}^2}
    \left(\frac{i f^2 t^3}{2^{17}}\right)^n \sum_{k=0}^{\infty}
    \frac{\Gamma(\frac{1}{2}+k)\,\Gamma\{4+6(n+k)\}}{\left(\Gamma\left\{\frac{5}{2}+3(n+k)\right\}\cdot k!\right)^2}
    \left(\frac{i f^2 t^3}{2^{11}}\right)^k
\end{equation}
[cf. (\ref{eq:propagator-III-5})]. We now apply the technique used
for (\ref{eq:propagator-III-6}) and (\ref{eq:propagator-III-7}),
which finally gives rise to
\begin{equation}\label{eq:propagator-IV-7}
    \mbox{I}_5(t)\, \approx\, -\frac{t^{3/2}}{3} \sqrt{\frac{1}{2\pi i}}\, \exp\left(-\frac{5 i f^2 t^3}{192}\right) J_0\left(\frac{f^2
    t^3}{64}\right)\, {}_{3}\hspace*{-.07cm}F_3\left(\left[\frac{1}{2},\frac{2}{3},\frac{4}{3}\right];
    \left[\frac{5}{6},\frac{7}{6},\frac{3}{2}\right]; \frac{i
    f^2t^3}{32}\right)\,,
\end{equation}
where
\begin{equation}\label{eq:propagator-IV-8}
    \frac{3}{32} \sum_{r=0}^{\infty}
    \frac{\Gamma\left(\frac{1}{2}+r\right)\,\Gamma(4+6r)}{\left\{\Gamma\left(\frac{5}{2}+3r\right)\cdot r!\right\}^2}
    \left(\frac{i f^2 t^3}{2^{11}}\right)^r\, =\,
    {}_{3}\hspace*{-.07cm}F_3\left(\left[\frac{1}{2},\frac{2}{3},\frac{4}{3}\right];
    \left[\frac{5}{6},\frac{7}{6},\frac{3}{2}\right]; \frac{i
    f^2t^3}{32}\right)
\end{equation}
[cf. (\ref{eq:propagator-III-7})]. Along the same line, we continue
to be able to obtain the next expressions in approximation as
\begin{eqnarray}
    \mbox{I}_6(t) &\approx& \frac{i t^2}{16}\, \exp\left(\frac{5 i f^2
    t^3}{192 i}\right) J_0\left(\frac{f^2
    t^3}{64}\right)\, {}_{4}\hspace*{-.07cm}F_4\left(\left[\frac{1}{2},\frac{5}{6},\frac{7}{6},\frac{3}{2}\right];
    \left[1,1,\frac{4}{3},\frac{5}{3}\right]; \frac{i
    f^2t^3}{32}\right)\\
    && \cdots\n\\
    \mbox{I}_{11}(t) &\approx& \frac{\{(2\pi i)^{-1}\, t^9\}^{\frac{1}{2}}}{3\cdot 5 \cdot 7\cdot 9 i}\, \exp\left(\frac{5 i f^2
    t^3}{192 i}\right) J_0\left(\frac{f^2
    t^3}{64}\right)\, {}_{4}\hspace*{-.07cm}F_4\left(\left[\frac{1}{2},\frac{5}{3},2,\frac{7}{3}\right];
    \left[1,\frac{11}{6},\frac{13}{6},\frac{5}{2}\right]; \frac{i
    f^2t^3}{32}\right)\n\\
    \mbox{I}_{12}(t) &\approx& -\frac{t^5}{2^6\cdot 5!}\, \exp\left(-\frac{5 i f^2
    t^3}{192}\right) J_0\left(\frac{f^2
    t^3}{64}\right)\, {}_{4}\hspace*{-.07cm}F_4\left(\left[\frac{1}{2},\frac{11}{6},\frac{13}{6},\frac{5}{2}\right];
    \left[1,2,\frac{7}{3},\frac{8}{3}\right]; \frac{i
    f^2t^3}{32}\right)\n\\
    && \cdots \;.\n
\end{eqnarray}
By induction it easily follows that
\begin{equation}\label{eq:semiclassial-n-1}
    \mbox{I}_n(t)\, \approx\, \mbox{I}_2(t)\cdot \frac{1}{\Gamma\left(\frac{n}{2}\right)}
    \left(\frac{t}{2i}\right)^{\frac{n}{2}-1}\,
    {}_{4}\hspace*{-.07cm}F_4\left(\left[\frac{1}{2},\frac{n-1}{6},\frac{n+1}{6},\frac{n+3}{6}\right];
    \left[1,\frac{n}{6},\frac{n+2}{6},\frac{n+4}{6}\right]; \frac{i f^2 t^3}{32}\right)
\end{equation}
where $n = 3, 4, \cdots$ [cf. Eqs. (\ref{eq:iterated1-1}) and
(\ref{eq:iterated2-1})]. Here the expression with $n=3$ is obviously
meant as the leading term of Eq. (\ref{eq:propagator19}) only. It is
also instructive to note that letting $f\to 0$, Eq.
(\ref{eq:semiclassial-n-1}) directly recovers the exact expression
of its field-free counterpart, explicitly given by
$(t^{n/2}-1)/\{(2i)^{n/2}\cdot \Gamma(n/2)\}$, obtained from
substitution of (\ref{eq:kernel2}) with $f=0$ into
(\ref{eq:propagator23}).

Now we are ready to have a closed expression of the quantity
$\Lambda(\tau')$ in approximation, accounting for the
ionization-recombination, as discussed after Eqs.
(\ref{eq:exponential_decay_ansatz}) and
(\ref{eq:propagator-expand-field1}). The substitution of
(\ref{eq:semiclassial-n-1}) into (\ref{eq:propagator-field1-1-0-0})
explicitly reveals that
\begin{eqnarray}\label{eq:propagator-field1-1}
    \Lambda_a(\tau') &\approx& \delta(\tau')\, +\, i K_f(\tau')\, +\,
    \frac{i}{2}\,\exp\left(-\frac{5 i f^2
    \tau'^3}{192}\right)\,J_0\left(\frac{f^2 \tau'^3}{64}\right) \left\{1\, +\,
    \sum_{n=3}^{\infty} \frac{1}{\Gamma\left(\frac{n}{2}\right)}
    \left(\frac{i \tau'}{2}\right)^{\frac{n}{2}-1}\, \times\right.\n\\
    && \left.{}_{4}\hspace*{-.07cm}F_4\left(\left[\frac{1}{2},\frac{n-1}{6},\frac{n+1}{6},\frac{n+3}{6}\right];
    \left[1,\frac{n}{6},\frac{n+2}{6},\frac{n+4}{6}\right]; \frac{i f^2
\tau'^3}{32}\right)\right\}\,.
\end{eqnarray}
This expression is useful especially in the weak-field limit, as
stated after Eq. (\ref{eq:semiclassial-n-1}). To see transparently
the strong-field behavior of $\Lambda_a(\tau')$, it may also be
desirable to rewrite (\ref{eq:propagator-field1-1}) as follows; we
first decompose the sum index $n$ as $6k+3, 6k+4, 6k+5, 6k+6, 6k+7$,
and $6k+8$, where $k = 0,1,\cdots$. Next we plug the expansion
${}_{4}\hspace*{-.07cm}F_4([\cdots];[\cdots];z) =
\sum_{l=0}^{\infty} (\cdots)\,z^l/l!$ into
(\ref{eq:propagator-field1-1}). With the aid of the sum identity
$\sum_{k=0}^{\infty} \sum_{l=0}^{\infty} g(k,l) =
\sum_{r=0}^{\infty} \sum_{k=0}^r g(k,r-k)$ where $r = k+l$, we can
unify and then simplify the two $\tau'$-dependencies, explicitly
given by $(\cdots\,\tau')^{n/2-1}$ and
${}_{4}\hspace*{-.07cm}F_4(\cdots\,\tau'^3)$ in the infinite sum
over $n$. This finally allows us to obtain $\Lambda_a(\tau')$ as
\begin{eqnarray}\label{eq:propagator-field2-1}
    && \delta(\tau')\, +\, i\,K_f(\tau')\, -\n\\
    && \mbox{I}_2(\tau')
    \left\{1\,+\,\left(\frac{\pi i \tau'}{2}\right)^{\frac{1}{2}}
    \sum_{j=0}^5
    \frac{\Gamma\left(\frac{j+2}{2}\right)}{\Gamma\left(\frac{j+3}{2}\right)}
    \left(\frac{i \tau'}{2}\right)^{\frac{j}{2}}
    \sum_{k=0}^{\infty}
    \frac{\left(4/f^2\right)^k}{\Gamma\left(1+\frac{j}{2}+3k\right)\,\Gamma\left(k+\frac{1}{2}\right)\,\left\{\Gamma(1-k)\right\}^2}\, \times\right.\n\\
    &&\left.{}_{5}\hspace*{-.07cm}F_5\left(\left[1,\frac{2+j}{6},\frac{4+j}{6},\frac{6+j}{6},\frac{1}{2}-k\right];
    \left[\frac{3+j}{6},\frac{5+j}{6},\frac{7+j}{6},1-k,1-k\right]; \frac{i
    f^2 \tau'^3}{32}\right)\right\}
\end{eqnarray}
with $f \to \infty$ [note that the sum over $j$ is just a finite
one]. Here the generalized hypergeometric function
${}_{5}\hspace*{-.07cm}F_5(\cdots\, i f^2 \tau'^3/32)$ can be
understood as a compact form of $\mbox{Re}(\cdots) +
i\,\mbox{Im}(\cdots)$, explicitly given by
\begin{eqnarray}\label{eq:hypergeometric-imaginary1}
    &&
    {}_{9}\hspace*{-.07cm}F_{10}\left(\left[1,\frac{1-2k}{4},\frac{3-2k}{4},\frac{2+j}{12},\frac{4+j}{12},\frac{6+j}{12},\frac{8+j}{12},\frac{10+j}{12},\frac{12+j}{12}\right];\right.\n\\
    && \left.\left[\frac{1-k}{2},\frac{1-k}{2},\frac{2-k}{2},\frac{2-k}{2},\frac{3+j}{12},\frac{5+j}{12},\frac{7+j}{12},\frac{9+j}{12},\frac{11+j}{12},\frac{13+j}{12}\right];
    -\frac{f^4 \tau'^6}{2^{12}}\right)\,+\n\\
    && i\,\frac{f^2 \tau'^3}{32} \left(\frac{1-2k}{2}\right) \left(\frac{2+j}{6}\right)
    \left(\frac{4+j}{6}\right) \left(\frac{6+j}{6}\right)\,\times\n\\
    && {}_{9}\hspace*{-.07cm}F_{10}\left(\left[1,\frac{3-2k}{4},\frac{5-2k}{4},\frac{8+j}{12},\frac{10+j}{12},\frac{12+j}{12},\frac{14+j}{12},\frac{16+j}{12},\frac{18+j}{12}\right];\right.\n\\
    && \left.\left[\frac{1-k}{2},\frac{1-k}{2},\frac{2-k}{2},\frac{2-k}{2},\frac{3+j}{12},\frac{5+j}{12},\frac{7+j}{12},\frac{9+j}{12},\frac{11+j}{12},\frac{13+j}{12}\right];
    -\frac{f^4 \tau'^6}{2^{12}}\right)\,.
\end{eqnarray}
And it is worthwhile to comment that although the function
${}_{5}\hspace*{-.07cm}F_5([\cdots]; [\cdots, 1-k,1-k]; \cdots)$
itself is not well-defined for $k = 1,2,\cdots$, the composite
expression ${}_{5}\hspace*{-.07cm}F_5([\cdots]; [\cdots, 1-k,1-k];
\cdots)/\{\Gamma(1-k)\}^2$, given in (\ref{eq:propagator-field2-1})
indeed, is mathematically undisputed. As a result, the substitution
of the quantity $\Lambda_a(\tau')$ in (\ref{eq:propagator-field1-1})
or (\ref{eq:propagator-field2-1}) into
(\ref{eq:propagator-expand-field1-0}) can directly yield an explicit
expression of the propagator
\begin{equation}\label{eq:propagator-expand-field1-0-0}
    {\mathcal K}_a(x,t|x',0)\, \approx\, K_f(x,x';t)\, +\, i t^2\int_0^1
    dy\,y\, K_f(x;t(1-y)) \int_0^1 dy'\, K_f(x';ty(1-y'))\cdot\Lambda_a(tyy')
\end{equation}
in terms of the generalized hypergeometric functions and an integral
which can straightforwardly be evaluated to any sufficient degree of
precision. Here we used $\tau' = \tau y'$ and $\tau = t y$. In fact,
it can easily be shown that this remaining double integral is beyond
the scope of its evaluation in terms of any generalized
hypergeometric functions in reasonably simple form. From
(\ref{eq:propagator-expand-field1}), we can also obtain the
corresponding wavefunction $\psi_a(x,t)$ for an arbitrary initial
state.

Comments deserve here. The wavefunction $\psi_a(x,t)$ evolved from
the initial bound state $\psi_b(x)$ must, by construction,
accommodate the ripples observed in the ionization probability
${\mathcal{P}}_f(t)$ in (\ref{eq:ionization_prob1}). To have this
feature, we took the leading term only from each integral
$\mbox{I}_n(t)$: In the strong-field limit, the first term
$\phi_f(x,t)$ on the right-hand side of Eq.
(\ref{eq:propagator-expand-field1}) for $\psi_a(x,t)$ is dominant to
the second term, given by a double integral, which has therefore
been neglected in the analytical approach in \cite{ELK88}. Now, the
result in (\ref{eq:propagator-expand-field1-0-0}) allows us to study
systematically the next-order terms to $\phi_f(x,t)$ (in the
intermediate-field regime), which are responsible indeed for the
ripples resulting from the scattering from the delta-potential well
at $x=0$. In the weak-field limit, on the other hand, the second
term (corresponding to the influence of the residual zero-range
potential) is dominant to the first term subjected to the external
field only, and so the quantity $\Lambda_a(\tau')$, and so
${\mathcal K}_a(0,t|0,0)$, is a highly critical factor to
determination of the entire time-evolution $\psi_a(x,t)$. In fact,
the bound-state amplitude ${\mathcal A}_{\psi_f}(t) = {\mathcal
A}_{\phi_f}(t) + {\mathcal A}_{\delta}(t)$ of the ionization
probability in Eqs.
(\ref{eq:ionization_prob1})-(\ref{eq:homogeneous_ionization1}) then
reduces to a compact form as
\begin{equation}\label{eq:second-term-of-ionization2}
    {\mathcal A}_{\delta}(t)\, \to\, i \int_0^t d\tau\, \int_0^{\tau} d\tau'\,
    \phi_f(0,t-\tau)\, \phi_f(0,\tau-\tau')\, \Lambda_a(\tau')
\end{equation}
[cf. (\ref{eq:homog_sol1}) for the closed expression of
$\phi_f(0,t)$]. Finally, we also remark that in actual numerical
evaluation of (\ref{eq:propagator-expand-field1-0-0}) we need to
introduce a (small) constant $c(t)$ in order to exactly fulfill the
normalization condition $\int_{-\infty}^{\infty} dx\,|\psi_a(x,t)|^2
= 1$ in such a way that $\psi_a(x,t) = \phi_f(x,t)\,+\,c(t) \times i
t^2 \int_0^1 dy \int_0^1 dy' (\cdots)$.

\section{Conclusion}\label{sec:conclusion}
In summary, we have investigated the time evolution of a particle
subjected to both a uniform electrostatic field and an attractive
one-dimensional delta-function potential. We have systematically
derived the propagator ${\mathcal K}_f(x,t|x',0)$ of this system, in
which its essential ingredient ${\mathcal K}_f(0,t|0,0)$, accounting
for the ionization-recombination in the bound-continuum transition,
is exactly expressed in terms of the multiple hypergeometric
functions. And then, as our central finding, we have obtained the
ingredient ${\mathcal K}_f(0,t|0,0)$ in (appropriate) approximation,
expressed in terms of the generalized hypergeometric functions being
much more transparent to physically interpret and much more
accessible in their numerical evaluation than the multiple
hypergeometric functions. It has been shown that this finding
provides a much better approximation scheme than the exponential
decay approximation, where no ionization-recombination dynamics can
be treated. In our approach, we have not applied the energy-time
Fourier nor the equivalent Laplace transform, which has, on the
other hand, been normally a starting point for analytical study of
this system in references, e.g., \cite{KLE94,ELK88,ELB88,LUD87}. We
think that our approach will therefore provide a useful framework
for analytical study of the time-dependent ionization process in a
delta-function potential well under an {\em oscillatory} electric
field as the next task, for which one cannot straightforwardly apply
the convolution theorem of the Laplace transform to the relevant
Lippmann-Schwinger integral equation.

\section*{Acknowledgments}
The author would like to thank G.J. Iafrate (NC State University),
Y. Wang, and L. Zhou for stimulating discussions. He also
appreciates support from G.J. Iafrate during his stay with NC State
University.

\begin{figure}[htb]
\centering\hspace*{-1cm}{\includegraphics[scale=0.7]{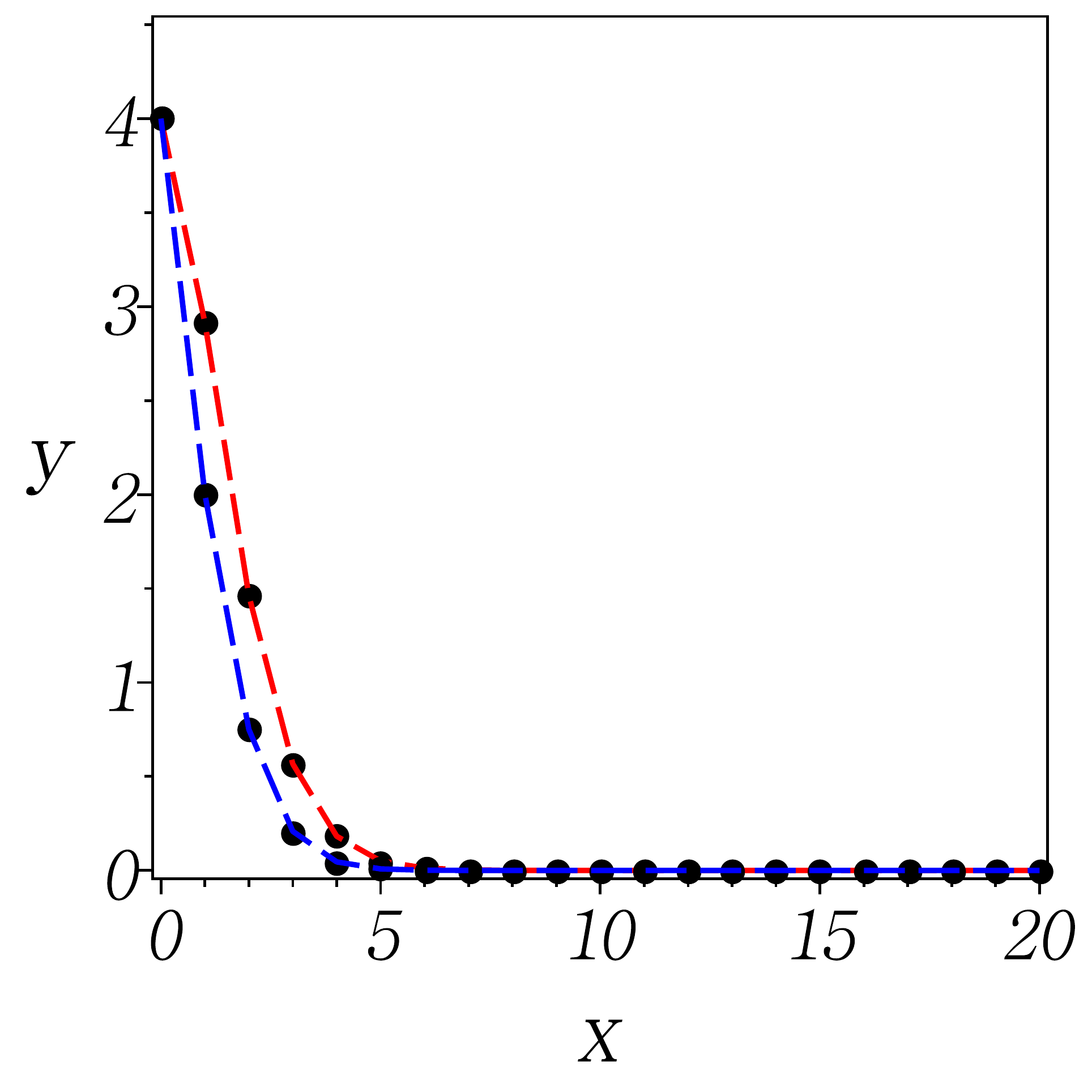}
\caption{\label{fig:fig1}}}
\end{figure}

\vspace*{2cm}
Fig.~\ref{fig:fig1}: (Color online) $y =
4\,\Gamma\left(\frac{1}{2}+r\right)/\{\sqrt{\pi}\,(r!)^2\}\,
{}_{4}\hspace*{-.07cm}F_3([\frac{1}{3},\frac{2}{3},-r,-r];
[\frac{5}{6},\frac{7}{6},\frac{1}{2}-r]; -1)$ versus index $r =
0,1,2,\cdots$ with an interpolated line (red dash). This precisely
represents the second summation over the index $n$ in Eq.
(\ref{eq:propagator-III-6}). In comparison, its approximation $y_a =
\Gamma(\frac{1}{2})\,\Gamma(2)\,\Gamma(\frac{1}{2}+r)/\{\Gamma(\frac{3}{2})\,\Gamma(1+r)\}^2$
with an interpolated line (blue dash), where $y > y_a$. This is
simply the leading term $n=0$ of the summation. As seen, $y \approx
y_a$ in a satisfactory manner.

\begin{thebibliography}{1}
%
\bibitem{YAM77} T. Yamabe, A. Tachibana, and H.J. Silverstone, Phys. Rev. A {\bf 16}, 877 (1977), and
references therein.
%
\bibitem{Raz03} M. Razavy, {\em Quantum Theory of Tunneling}
(World Scientific, Singapore, 2003).
%
\bibitem{JOA86} C. Joachim and J.P. Launay, Chem. Phys. {\bf 109}, 93 (1986).
%
\bibitem{JON89} M. Jonson, Phys. Rev. B {\bf 39}, 5924 (1989).
%
\bibitem{GEL77} S. Geltman, J. Phys. B {\bf 10}, 831 (1977) and J. Phys. B {\bf 11}, 3323 (1978).
%
\bibitem{LUD87} A. Ludviksson, J. Phys. A {\bf 20}, 4733 (1987).
%
\bibitem{ELK88} W. Elberfeld and M. Kleber, Z. Phys. B {\bf 73}, 23 (1988).
%
\bibitem{SUS90} S.M. Susskind, S.C. Cowley, and E.J. Valeo, Phys. Rev. A {\bf 42}, 3090 (1990).
%
\bibitem{KLE94} M. Kleber, Phys. Rep. {\bf 236}, 331 (1994).
%
\bibitem{ENG95} B.-G. Englert, Lett. Math. Phys. {\bf 34}, 239 (1995); here, an ``analytical
expression'' of the ionization probability ${\mathcal P}_f(t)$ was
derived, which is, however, just a complicated formal expression in
terms of some differential operator [cf. Eqs. (29) and (35)
thereof].
%
\bibitem{ROK00} A. Rokhlenko and J.L. Lebowitz, J. Math. Phys. {\bf 41}, 3511
(2000); here, the time-dependent wavefunction was expanded in terms
of a complete set of eigenfunctions of the field-free Hamitonian
$\hat{\mathcal H}_0 = \hat{p}^2/2m - V_0\,\delta(x)$ to study the
system of a single delta-function potential well which is subject to
{\em time-dependent} variations of the binding strength, but with no
external field.
%
\bibitem{ELB87} W. Elberfeld and M. Kleber, {\em Ionization and tunneling in a strong electric
field}, in {\em Physics of Strong Fields}, edited by W. Greiner
(Plenum, New York, 1987) p. 465.
%
\bibitem{REI85} W.P. Reinhardt, {\em Time-dependent approach to spectra in external fields},
in {\em Atomic Excitation and Recombination in External Fields},
edited by M.H. Nayfeh and C.W. Clark (Gordon and Breach, New York,
1985).
%
\bibitem{GRI05} D.J. Griffiths, {\em Introduction to Quantum Mechanics}, 2nd ed. (Pearson
Prentice Hall, New Jersey, 2005).
%
\bibitem{DAM75} W.C. Damert, Am. J. Phys. {\bf 43}(6), 531 (1975).
%
\bibitem{VAL04} O. Vall\'{e}e and M. Soares, {\em Airy Functions and Applications to
Physics}, 2nd ed. (World Scientific, Singapore, 2010).
%
\bibitem{WEI95} S. Weinberg, {\em The Quantum Theory of Fields} (Cambridge
University Press, 1995).
%
\bibitem{GRO98} C. Grosche C and F. Steiner, {\em Handbook of Feynman path integrals} (Springer, New York, 1998).
%
\bibitem{MOS52} M. Moshinsky, Phys. Rev. {\bf 88}, 625 (1952).
%
\bibitem{ABS74} M. Abramowitz and I. Stegun, {\em Handbook of Mathematical Functions
with Formulas, Graphs, and Mathematical Tables} (Dover, New York,
1974).
%
\bibitem{SCH84} R.R. Schlicher, W. Becker, J. Bergou, and M.O.
Scully, {\em Interaction Hamiltonian in Quantum Optics or
$\hat{p}\,A$ vs. $E\,\hat{r}$ Revisited}, in {\em Quantum
Electrodynamics and Quantum Optics}, edited by A.O. Barut (Plenum,
New York, 1984) p. 405.
%
%
\bibitem{BAU85} D. Bauch, Nuovo Cimento B {\bf 85}, 118 (1985).
%
\bibitem{ELB88} W. Elberfeld and M. Kleber, Am. J. Phys. {\bf 56}(2), 154 (1988).
%
\bibitem{EXT76} H. Exton, {\em Multiple Hypergeometric Functions and Applications} (Halsted/Wiley, New York, 1976).
%
\bibitem{SRI85} H.M. Srivastava and P.W. Karlsson, {\em Multiple Gaussian
Hypergeometric Series} (Halsted/Wiley, New York, 1985).
%
\bibitem{SRI87} H.M. Srivastava, J. Phys. A {\bf 20}, 847 (1987).
%
\bibitem{SOM58} D.M.Y. Sommerville, {\em An introduction to the geometry of n
dimensions} (Dover, New York, 1958).
%
\bibitem{KIM12} Eqs. (\ref{eq:propagator14}),
(\ref{eq:propagator17}) and (\ref{eq:propagator18}) as well as all
following expressions in terms of the hypergeometric functions have
been verified with the help of Maple programs.
%
\end{thebibliography}
\end{document}